\documentclass[aps,prb,reprint,showpacs,twocolumn,superscriptaddress,floatfix]{revtex4-1}
\usepackage{graphicx}
\usepackage{amssymb}
\usepackage{amsmath}
\usepackage[FIGTOPCAP,nooneline]{subfigure}
\usepackage{overpic}

\begin{document}

\title{Secondary "Smile"-gap in the density of states of a diffusive Josephson junction for a wide range of contact types }

\author{J. Reutlinger}

\affiliation{Fachbereich Physik, Universit\"at Konstanz, D-78457 Konstanz, Germany}

\author{L. Glazman}

\affiliation{Department of Physics, Yale University, New Haven CT 06511-8499, USA}

\author{Yu. V. Nazarov}

\affiliation{Kavli Institute of Nanoscience Delft, Delft University of Technology, 2628 CJ Delft, The Netherlands}

\author{W. Belzig}

\affiliation{Fachbereich Physik, Universit\"at Konstanz, D-78457 Konstanz, Germany}

\email[]{Wolfgang.Belzig@uni.kn}


\date{\today}

\begin{abstract}

The superconducting proximity effect leads to strong modifications of the local density of states in diffusive or chaotic cavity Josephson junctions, which displays a phase-dependent energy gap around the Fermi energy. The so-called minigap of the order of the Thouless energy $E_{\mathrm{Th}}$ is related to the inverse dwell time in the diffusive region in the limit $E_{\mathrm{Th}}\ll\Delta$, where $\Delta$ is the superconducting energy gap. In the opposite limit of a large Thouless energy $E_{\mathrm{Th}}\gg\Delta$, a small new feature has recently attracted attention, namely, the appearance of a further secondary gap, which is around two orders of magnitude smaller compared to the usual superconducting gap. It appears in a chaotic cavity just below the superconducting gap edge $\Delta$ and vanishes for some value of the phase difference between the superconductors. We extend previous theory restricted to a normal cavity connected to two superconductors through ballistic contacts to a wider range of contact types. We show that the existence of the secondary gap is not limited to ballistic contacts, but is a more general property of such systems. Furthermore, we derive a criterion which directly relates the existence of a secondary gap to the presence of small transmission eigenvalues of the contacts. For generic continuous distributions of transmission eigenvalues of the contacts, no secondary gap exists, although we observe a singular behavior of the density of states at $\Delta$. Finally, we provide a simple one-dimensional scattering model which is able to explain the characteristic "smile" shape of the secondary gap.

\end{abstract}


\maketitle


\section{Introduction}

One of the most striking impacts of a contact with a superconductor (S) onto a small piece of normal metal (N) is the modification of the local density of states (LDOS). This effect, known as the superconducting proximity effect, is related to the induction of superconducting correlations resulting in a finite value of the pair amplitude $\sim\langle \hat \Psi_\uparrow (\vec r) \hat \Psi_\downarrow (\vec r) \rangle$ on the normal side \cite{deutscher:69}. In the absence of phonon mediated attraction between electrons on the normal side, decoherence between electronlike and holelike amplitudes leads to an exponential decay of the pair amplitude with distance from the contact, with a  characteristic length scale exceeding the superconducting coherence length. 

Modification of the LDOS on both sides of the contact strongly depends on the scattering properties of the contacts (described in terms of transmission eigenvalues) and the properties of the normal region (geometry, size, and impurity concentration). In the case of diffusive systems, it was predicted theoretically that the LDOS can even be fully suppressed in a specific energy range around the Fermi energy which is known as the minigap \cite{mcmillan:68}. The minigap width is of the order of the inverse dwell time in the normal structure, which is given by the Thouless energy $E_{\textrm{Th}}=(G_\Sigma/G_Q) \delta_s$, where $\delta_s$ denotes the mean level spacing of the normal region and $G_\Sigma \gg G_Q$ is the total conductance of the structure which is assumed large compared to the conductance quantum $G_Q=e^2/\pi \hbar$. In the decades after its discovery it has in detail been studied theoretically \cite{golubov:88,beenakker:92,belzig:96}. The development of more elaborate experimental techniques with high spatial resolution made variations of the LDOS in this energy range accessible to experiments \cite{gueron:96,scheer:01,moussy:07,mourik:12, churchill:13, garcia:13, cherkez:14}, which was found to be in agreement with theoretical calculations to a high degree \cite{leseur:08,pillet:10,wolz:11}.

Much interest was concentrated on systems built up of a finite normal region sandwiched between two superconductors: a Josephson junction \cite{josephson:62}. In such systems, another parameter, i.e., the phase difference between the superconducting order parameters comes into play and leads to a phase-dependent minigap \cite{belzig:99,golubov:04}. Classical ballistic systems \cite{lodder:98} were investigated as well as diffusive systems \cite{belzig:96} and the crossover between both \cite{pilgram:00}. It turns out that not only diffusive systems exhibit a minigap, but also ballistic systems with a chaotic classical motion \cite{melsen:97,lodder:98,VavilovLarkin:03, beenakker:05,  kuipers:10, kuipers:11}.

At this point, one might think that such structures are sufficiently explored and all relevant properties are understood. However, recently Levchenko reported the finding of a dip in the LDOS close to the gap edge $\Delta$ for short diffusive Josephson junctions with ideal contacts \cite{levchenko:08}. Actually, this dip was already seen in former publications \cite{golubov:96,belzig:96, wilhelm:00, heikkilae:02,  bezuglyi:05, hammer:07}, however, no special attention was paid to it. In a previous work \cite{reutlinger:14} we found the peculiar result that the suppression of the LDOS at $\Delta$ is not limited to a dip, but a secondary gap of finite width appears for a diffusive system or chaotic cavity with the normal region connected through ballistic contacts to the superconductors. This secondary gap has a finite width as a function of the superconducting phase difference $\varphi$ symmetrically around zero and closes with the characteristic shape of a "smile". It is situated directly below the superconducting gap edge $\Delta$ for large $E_{\textrm{Th}}\gtrsim\Delta$. For decreasing $E_{\textrm{Th}}$ the upper edge of the secondary gap detaches from $\Delta$ and the gap vanishes completely below a critical value of $E_{\textrm{Th}}$. We have furthermore shown that the secondary gap is robust against asymmetries in the setup, comprising a difference in ballistic couplings or a weak spatial dependence.

In this work, we investigate a wide range of possible nonballistic contacts and show that the secondary "smile"-gap is not only an exotic feature which appears for ballistic contacts, but is a more general property of short diffusive or chaotic Josephson systems. Using quasiclassical Green's functions in the form of the quantum circuit theory, we begin by generalizing our ballistic calculations to contacts with constant transmission eigenvalues $<1$, for which we calculate the density of states as a function of $E_{\textrm{Th}}$ and $\varphi$. We find a secondary gap which scales for large $E_{\textrm{Th}}$ like in the ballistic case. In a specific example of contacts described by different constant transmission eigenvalues we show that one is not limited to a single secondary gap, but this gap can be split up into multiple subgaps. Numerical considerations of continuous transmission distributions (diffusive, dirty, double ballistic contacts) suggest that the secondary gap below the superconducting gap edge vanishes if the contacts include channels with close-to-zero transmission coefficients. We prove this conjecture by an analytical calculation. By considering asymmetric setups with a tunnel contact on one and a ballistic contact on the other side we show however, that in this case a secondary gap can exist at slightly smaller energies. By considering a 3-node system we show that although the LDOS varies at different nodes the secondary gap appears either in all nodes or in none of them. It should as well be observable in the integrated DOS of the normal part. Finally, we provide a simple one-dimensional (1D) model in order to describe transmission through the normal region, which is able to explain the "smile" shape of the secondary gap.

\section{Model}

\begin{figure}[t] 
\vspace{5mm}
\begin{overpic}[width=0.95\columnwidth,angle=0]{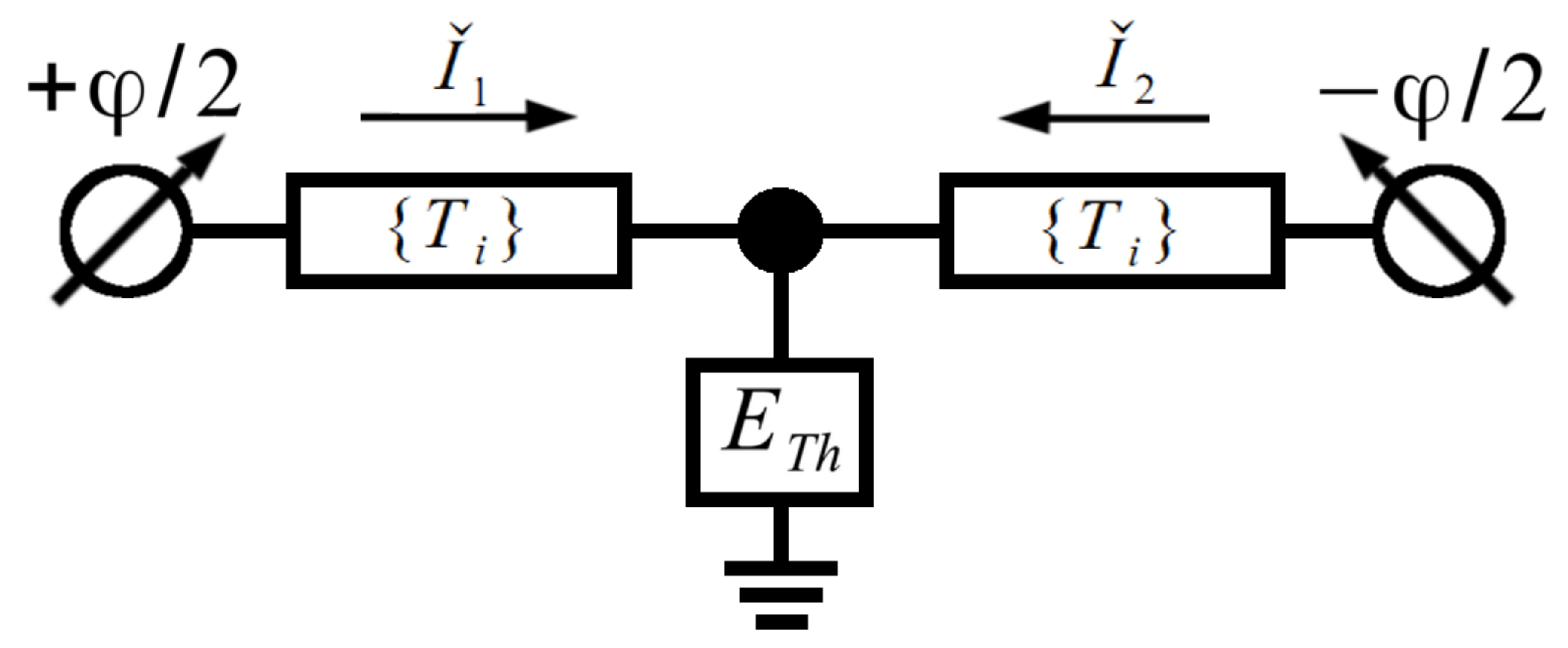}
\put(3,43){\makebox(0,3){$(a)$}}
\end{overpic}
\begin{overpic}[width=0.95\columnwidth,angle=0]{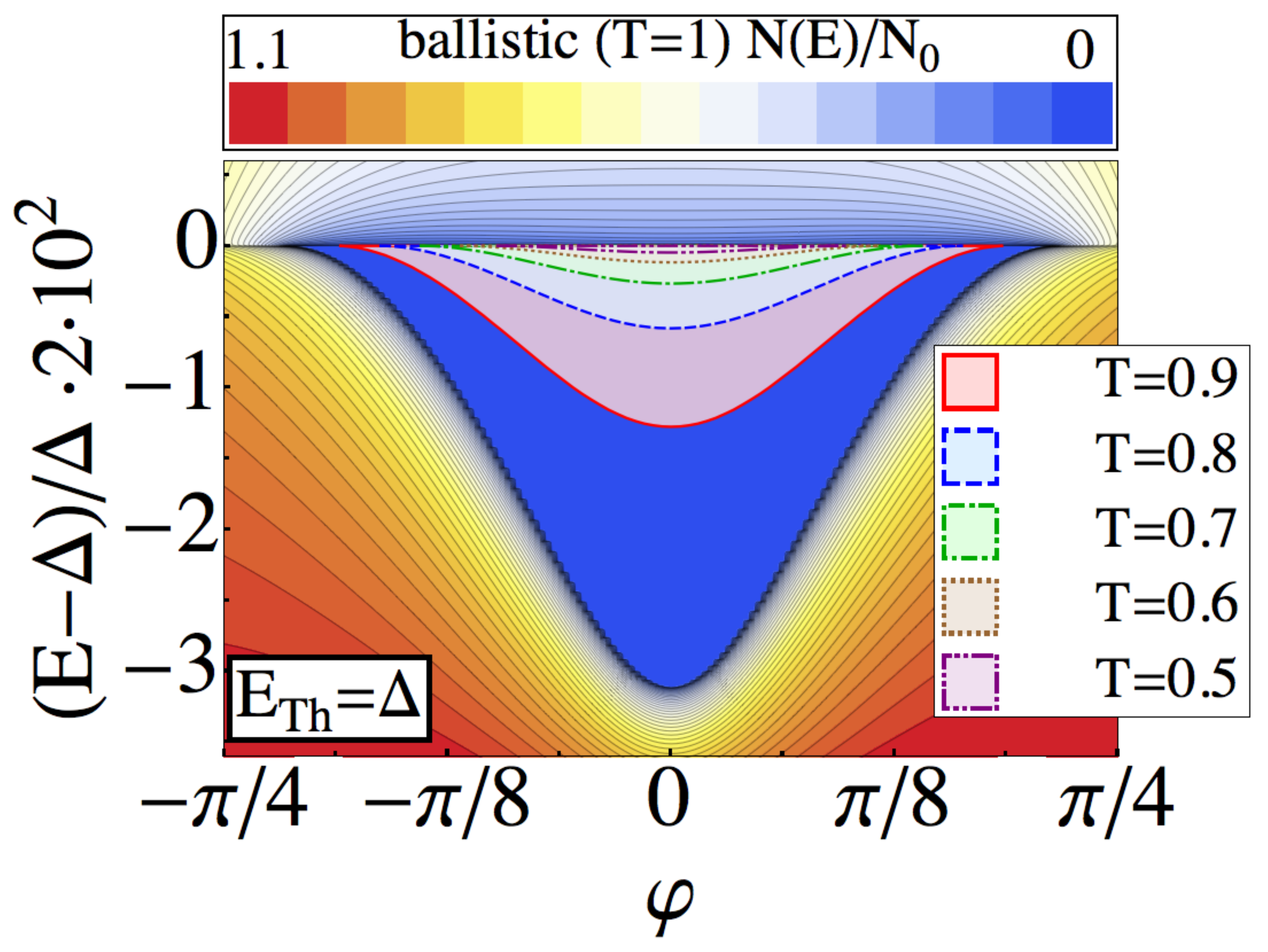}
\put(3,75){\makebox(0,3){$(b)$}}
\end{overpic}
\caption{\label{fig:1}
(a) Schematic representation of the investigated structure in discretized form. Both superconductors have equal energy gaps $\Delta$ and a relative phase difference $\varphi$. Information on the size of the normal region is contained in the Thouless energy $E_{\textrm{Th}}=\hbar/\tau$, $\tau$ being the average dwell time in the normal part. In discretized form, the Green's function in the normal node is determined by requiring current conservation under consideration of matrix currents to both superconductors as well as a leakage current related to $E_{\textrm{Th}}$. The transmission properties of the contacts are described by a set of transmission eigenvalues $\lbrace T_\textrm{n}^{\textrm{i}} \rbrace $,which can in general be different for the two leads $i$.\\
(b) Contour of the secondary gap for different constant transmissions $T=0.9$ (red), $T=0.8$ (blue), $T=0.7$ (green), $T=0.6$ (brown) and $T=0.5$ (purple) compared to the LDOS for ballistic contacts.}
\end{figure}

In order to calculate the LDOS in the normal region, we make use of the retarded Green's function in the quasiclassical approximation. In the diffusive or dirty limit, the angle-averaged Green's functions are described by the nonlinear diffusive Usadel equation which has the form of a continuity equation for coherence functions including the leakage of coherence due to the finite energy difference between electrons and holes. Since the spatial dependence of the Green's function is not important for our needs (for more details see Sec. III. F), we can solve the problem by applying the so-called quantum circuit theory \cite{nazarov:94, qt, nazarov:99}. We can discretize the system and reduce the equations to an algebraic problem. A sketch of the investigated system is shown in Fig.~\ref{fig:1}. The superconductors have equal energy gaps $\Delta$, however in general the phases of the order parameters can be different. Since the global phase is of no significance, only the phase difference $\varphi$ enters our calculation and we can assign the phase $\pm\varphi/2$ to the left and right superconductors, respectively. The Green's function in the normal node ${\hat G}_c$ is determined by the constraint of matrix current conservation, including the currents to the two superconductors ${\hat I}_{\textrm{ic}}$ ($i$ being the index denoting left and right lead) as well as the leakage current related to the volume of the normal region through $E_{\textrm{Th}}$:
\begin{equation}
	\label{eq:curr_cons}
 	\hat I_{1c}+\hat I_{2c} +iG_{\Sigma}\frac{E}{E_{\textrm{Th}}}[\hat\tau_3,\hat G_c(E)]=0\,.
\end{equation}
The scattering properties of the contacts are contained in the expressions for the matrix currents \cite{nazarov:99}
\begin{equation}
\label{eq:curr}
{\hat I}_{ic}= 2 G_Q  \sum_n \frac {T_{n}^{i} (\hat G_c\hat G_{i}-\hat G_{i}\hat G_c)}{4+T_{n}^{i}(\hat G_c\hat G_{i}+\hat G_{i}\hat G_c-2)}
\end{equation}
in terms of a set of transmission eigenvalues $\lbrace T_n^{i} \rbrace$. In general the transmission eigenvalues can be different on both sides. For continuous transmission distributions $\rho_{\textrm{i}}(T)$ the sums must be replaced by integrals over the particular distributions. The Green's functions in the leads are those of a bulk superconductor, given by $\hat G_{1,2}=c\hat\tau_3+is[\hat\tau_1\cos(\varphi/2)\pm\hat\tau_2\sin(\varphi/2)]$ with the spectral functions $c$ and $s$ being given by $c=\sqrt{1+s^2}=E/\sqrt{E^2-\Delta^2}$ for $E>\Delta$ and by 
$c=\sqrt{1+s^2}=-iE/\sqrt{\Delta^2-E^2}$ for $E<\Delta$, $\hat\tau_i$ being the Pauli matrices in Nambu space of electrons and holes. In the normal node the Green's function can be parametrized as $\hat G_c=g\hat\tau_3+if[\hat\tau_1\cos(\phi/2)-\hat\tau_2\sin(\phi/2)]$. $g$ and $f$ are related via the normalization condition for quasi-classical Green's functions $\hat G_c^2=1$, which is equivalent to $g^2-f^2=1$.  In the general case with different contacts on both sides, this corresponds to the solution of two equations for two complex variables. Expanding (\ref{eq:curr_cons}) in Pauli matrices and comparing the coefficients provides two independent equations 


\begin{widetext}
\begin{eqnarray}
\label{eq:1}
	2i\frac{E}{E_{\textrm{Th}}} f \cos(\phi)+\left[ g s \cos(\varphi/2)-c f \cos(\phi) \right ] 
	\left[ \frac{X_1^{-1}}{1+G_2/G_1}+\frac{X_2^{-1}}{1+G_1/G_2} \right] & = & 0\\
\label{eq:2}
	-2i\frac{E}{E_{\textrm{Th}}} f \sin(\phi)
	+\left[ -g s \sin(\varphi/2)+c f \sin(\phi) \right ] \frac{X_1^{-1}}{1+G_2/G_1}
	+\left[ g s \sin(\varphi/2)+c f \sin(\phi) \right ] \frac{X_2^{-1}}{1+G_1/G_2} &=&0.
\end{eqnarray}
\end{widetext}

Note that $f$ and $g$ as well as $\phi$ are complex valued in general. All information on the contacts is contained in the characteristic functions $X_i^{-1}$ given by 

\begin{equation}
\label{eq:characteristic_funct}
	X_i^{-1} = \frac{G_Q}{G_i}
	\sum_n \frac{T_{n}^{i}}{1+T_{n}^{i}(a_i-1)/2},
\end{equation}
with $a_{1/2}=-f s \cos(\phi \mp \varphi/2)+cg$ and $G_{i}=G_{\textrm{Q}} \sum_n T_{n}^{i}$ being the conductance of the particular side. Again, for a continuous transmission distribution, the sums must be replaced by integrals over the particular distributions $\rho_{i}(T)$. For a symmetric setup $X_1=X_2=X$ and $\phi=0$. Equation (\ref{eq:2}) becomes trivial and only one equation in one complex variable remains. From Eq. (\ref{eq:1}) we find
\begin{equation}
\label{eq:3}
2iE/E_{\textrm{Th}} f+\left ( g s \cos(\varphi/2)-c f \right ) X^{-1}=0.
\end{equation}
The density of states $N(E)$ is finally obtained from $\hat{G}_c$ through $N(E)/N_0=\mathrm{Re}\{\mathrm{Tr}\hat \tau_3\hat G_c(E)\}/2=\mathrm{Re}\{g\}$, $N_0$ being the density of states at the Fermi energy of the normal state.

\section{results}

In previous analysis \cite{reutlinger:14} this setup was investigated for ballistic contacts  with all $T_i=1$. It turned out that the secondary "smile"-gap which appears in the symmetric case is stable under asymmetries $G_1/G_2\ne 1$. For an asymmetric setup, two further gaps, complementary to the usual minigap and the "smile"-gap, appear symmetrically around $\varphi=\pi$. In this work, we want to extend these calculations and consider a wider range of contact types, corresponding to a wider range of characteristic functions $X_i^{-1}$, either described by discrete transmission eigenvalues or by continuous distributions $\rho_{i}(T)$. The idea of this work is to investigate the stability of the secondary gap under deviation from the ballistic limit. Especially, we want to determine which contact properties define the existence of the secondary gap, since it is known that for tunnel contacts no secondary gap is found. For this reason, we consider symmetric as well as asymmetric setups in the intermediate regime between the tunnel and ballistic limits.

\subsection{Constant transmission $T< 1$}

\begin{figure}[t] 
\vspace{5mm}
\begin{overpic}[width=0.95\columnwidth,angle=0]{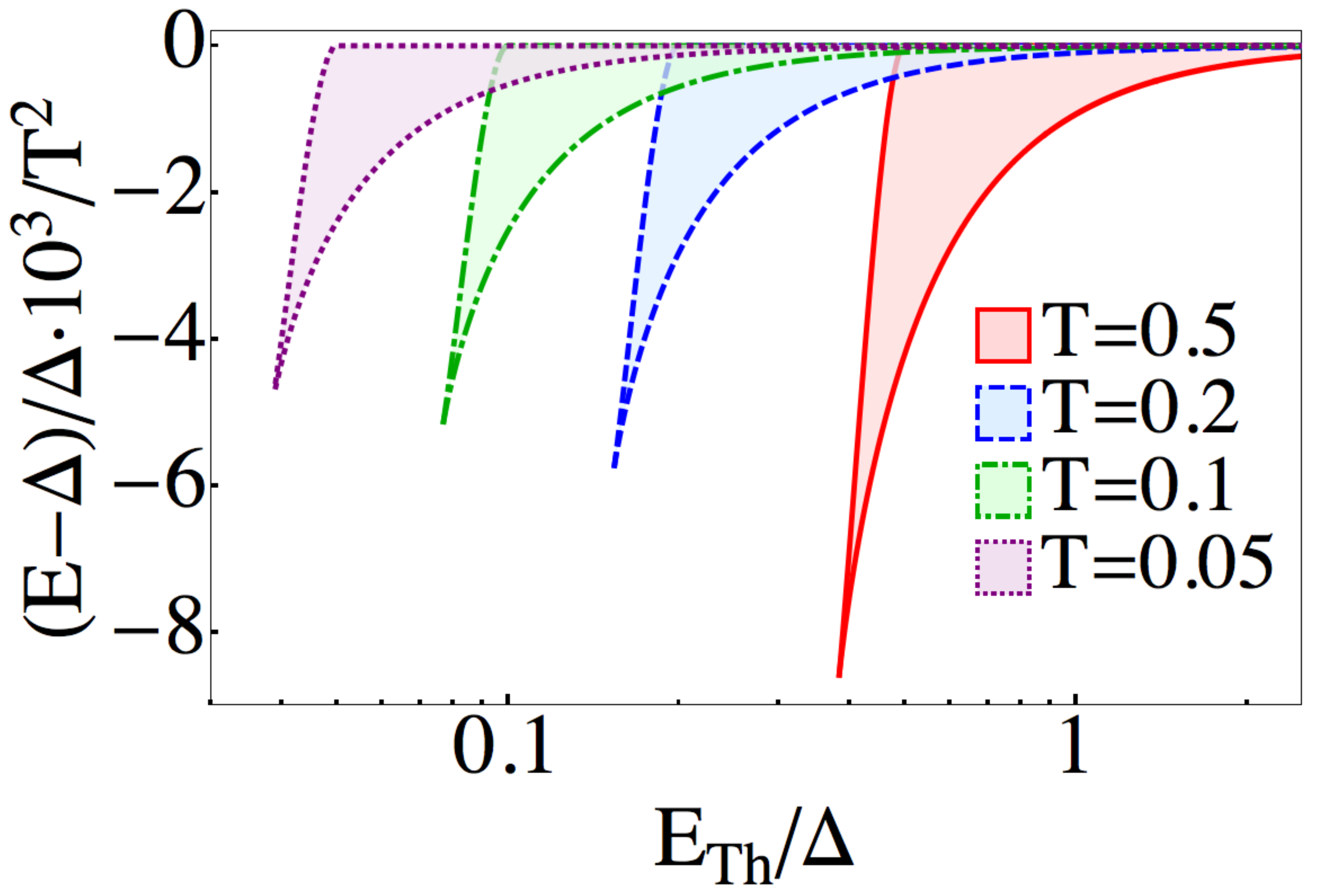}
\put(3,66){\makebox(0,3){$(a)$}}
\end{overpic}
\begin{overpic}[width=0.95\columnwidth,angle=0]{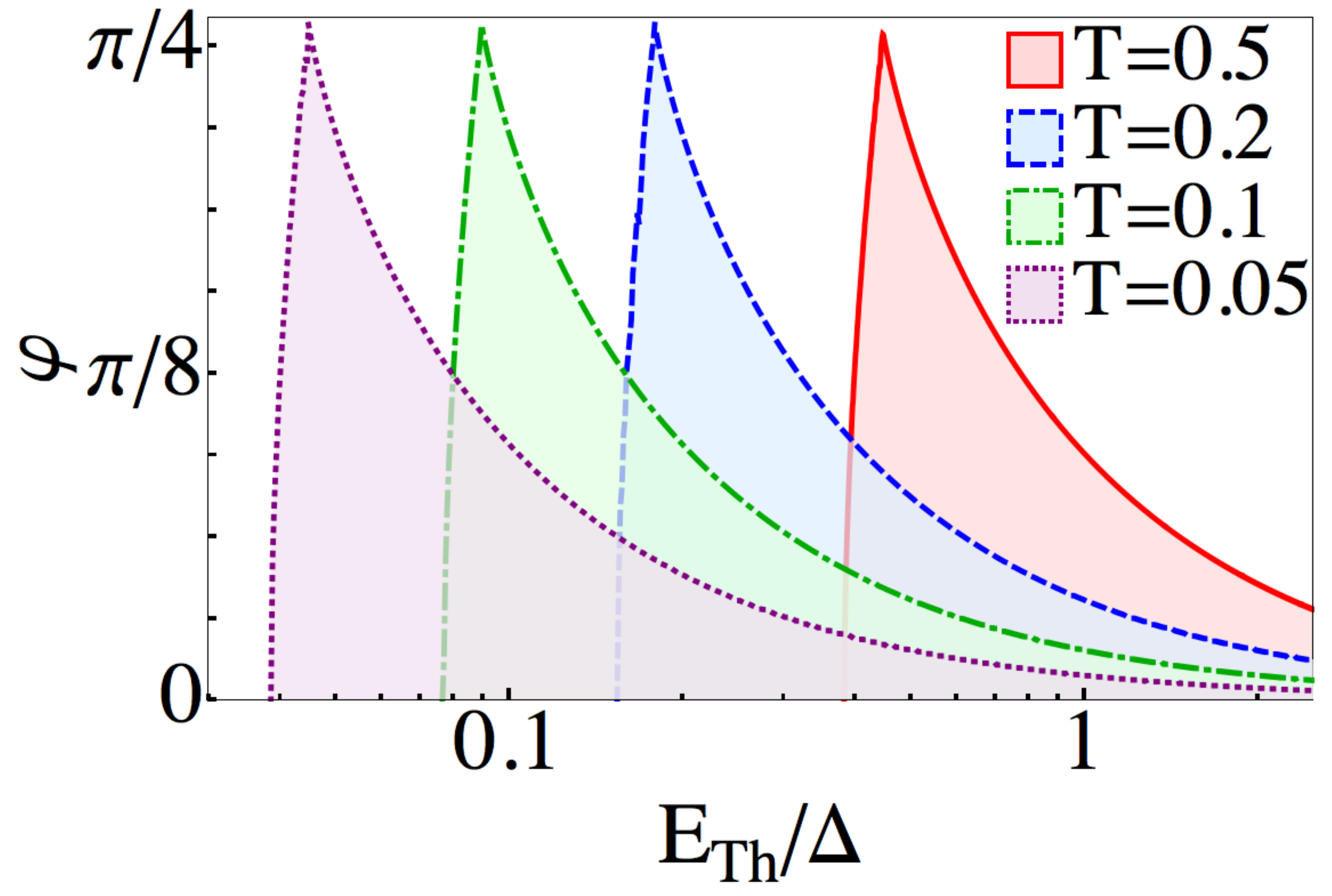}
\put(3,66){\makebox(0,3){$(b)$}}
\end{overpic}
\caption{\label{fig:2} (a) For constant transmission eigenvalues $T<1$, a secondary gaps appears below $\Delta$ similar to the ballistic case. For large Thouless energies, the upper edge is attached to $\Delta$, at  $E_{\textrm{Th,det}}=T \Delta$ it detaches from $\Delta$ and approaches the lower edge until the gap disappears at some critical value which seems to scale linearly with $T$ as well.  The maximum width of the gap decreases with decreasing $T$. The plot shows the numerical results for the critical parameters $E^{\textrm{upp}}_{\textrm{c}}$ and $E^{\textrm{low}}_{\textrm{c}}$  at $\varphi=0$ for different values of the constant transmission eigenvalue $T$. 
(b) Dependence of the critical phase $\varphi_c$ on $E_{\textrm{Th}}$. The maximum of the critical phase does not change with $T$. However the $E_{\textrm{Th}}$ dependence seems to scale linearly with $T$ and is shifted to smaller Thouless energies for decreasing $T$. }
\end{figure}

A natural generalization of the ballistic contact is to stick to constant transmission eigenvalues, however, to allow for $T<1$. As $T$ approaches 0, the secondary gap is expected to disappear and the tunnel result for the LDOS should be reproduced. We begin by considering symmetric contacts and thus solve Eq. (\ref{eq:3}) with the characteristic function
\begin{displaymath}
X=1+T/2(a-1).
\end{displaymath}
We find a secondary gap in the LDOS similar to the ballistic result, which survives even for small but finite $T$. The numerical results for the critical phase $\varphi_c$, for which the gap closes, as well as for the upper and lower gap edges $E^{\textrm{upp}}_{\textrm{c}}$ and $E^{\textrm{low}}_{\textrm{c}}$ at $\varphi=0$, are shown in Fig.~\ref{fig:2}. The colored regions denote the gap. Above a special value $E_{\textrm{Th,det}}$, which scales linearly with $T$ and is given by $E_{\textrm{Th,det}}=T\Delta$, the upper gap edge is fixed to $\Delta$ and the lower edge approaches $\Delta$ for increasing $E_{\textrm{Th}}$ following a power law. The linear scaling of $E_{\textrm{Th,det}}$ with $T$ follows from Eq. (\ref{eq:3}) for $E=\Delta$ and $\varphi=0$. The dependence of the lower gap edge $E^{\textrm{low}}_{\textrm{c}}$ on $E_{\textrm{Th}}$ for $E_{\textrm{Th}}\gg \Delta$ is derived in the following. 

Below $E_{\textrm{Th,det}}$, the upper edge is detached from $\Delta$ and approaches the lower edge until the gap disappears at some critical value of $E_{\textrm{Th}}$ which as well seems to scale linearly with $T$. The maximum of the critical phase at which the secondary gap closes [Fig.~\ref{fig:2} (b)] does not depend on $T$, however, it is shifted to smaller Thouless energies with decreasing $T$. The dependence of the critical phase on $E_{\textrm{Th}}$ seems to scale linearly with $T$. In the limit $T \to 0$, the gap disappears and the tunnel result without secondary gap is reproduced. However, for each finite value of $T$ the secondary gap exists, if the Thouless energy is made large enough. To get further insight to the analytic properties we linearize Eq. (\ref{eq:3}) in the energy range below $E=\Delta$ and around $\varphi=0$ in the limit $E_{\textrm{Th}}\gg \Delta$. We find
\begin{equation}
\label{eq:4}
\begin{split}
 0 = & \frac{1}{2 g^2}+\left( \delta-\varphi^2/8 \right)-\frac{2 \Delta}{E_{\textrm{Th}}}\sqrt{2\delta} \\
 & \times \left\{1+\frac{T}{2} \left[\frac {-ig}{\sqrt{2\delta}} \left (\frac{1}{2g^2}- \delta+ \frac{\varphi^2}{8} \right)-1\right]\right\} ,
\end{split}
\end{equation}
with $\delta=(\Delta-E)/\Delta$ being the dimensionless energy relative to $\Delta$.
\begin{figure}[t] 
\vspace{5mm}
\begin{overpic}[width=0.95\columnwidth,angle=0]{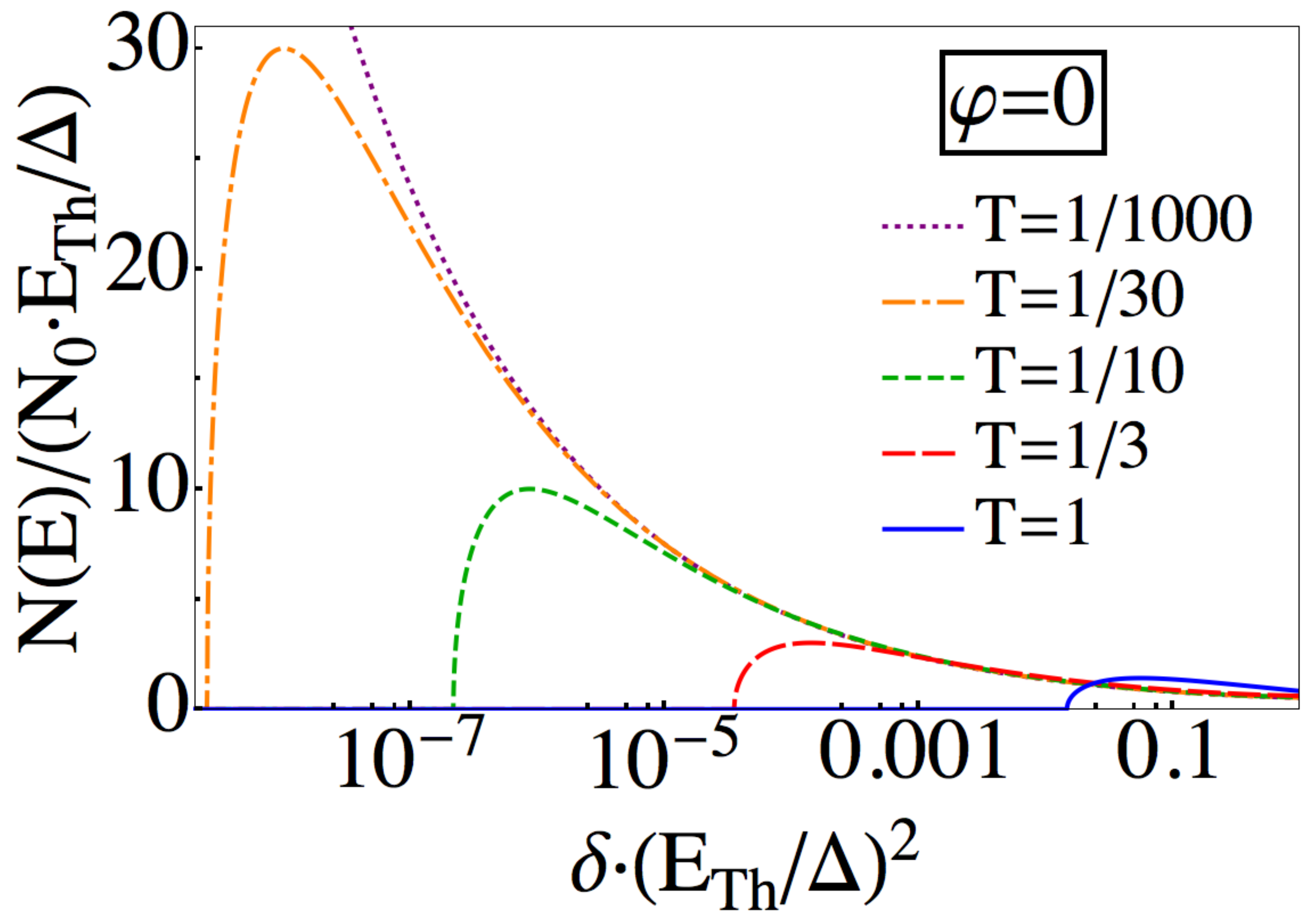}
\put(3,66){\makebox(0,3){$(a)$}}
\end{overpic}
\begin{overpic}[width=0.95\columnwidth,angle=0]{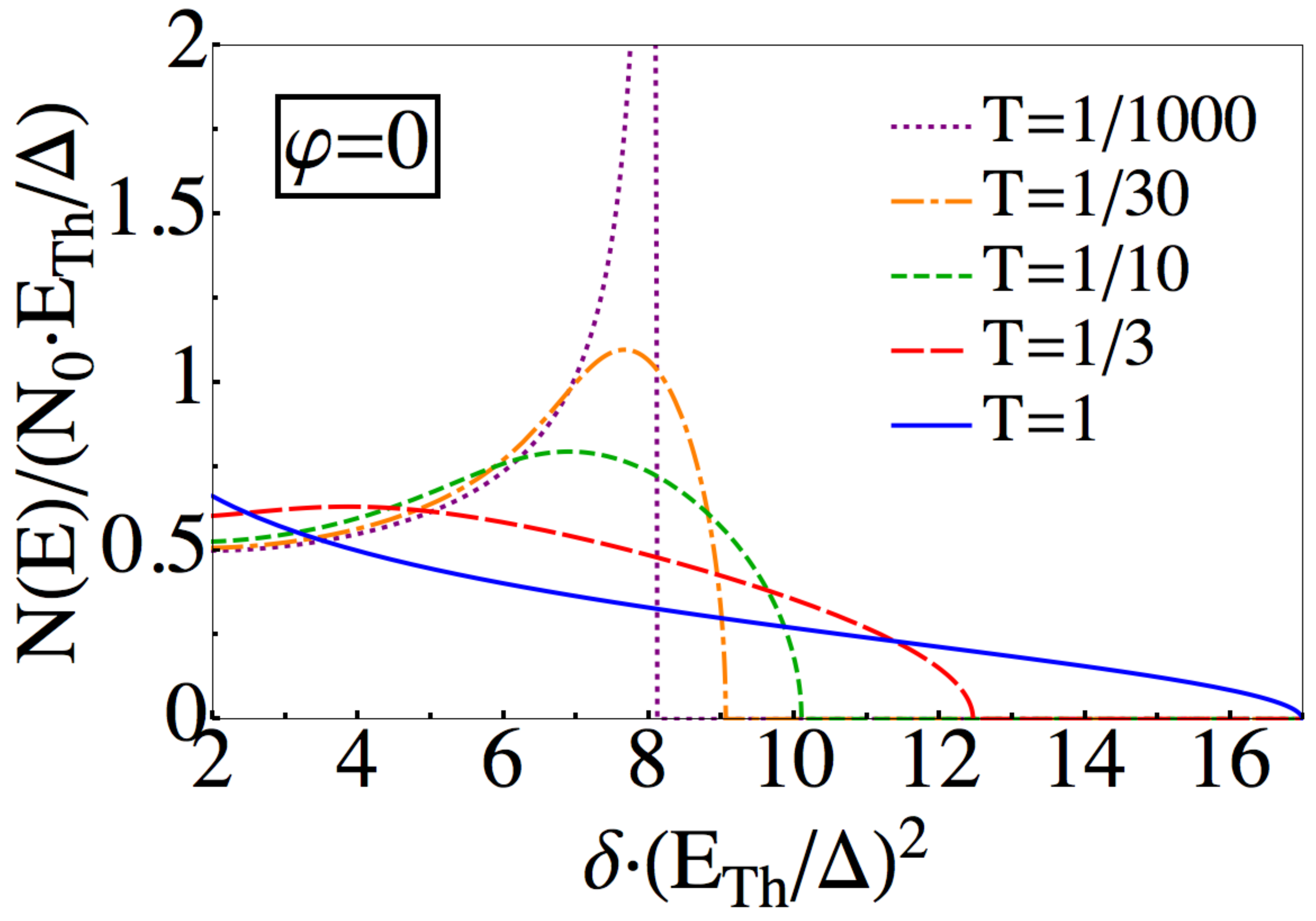}
\put(3,66){\makebox(0,3){$(b)$}}
\end{overpic}
\caption{\label{fig:3} Solution of the linearized Eq. (\ref{eq:4}) being valid for large $E_{\textrm{Th}}$ in the parameter range of interest [i.e. $\delta=(\Delta-E)/\Delta \ll 1$, $\varphi \ll 1$]. For finite $T$, a secondary gap appears, however, with decreasing $T$ the gap shrinks and the LDOS approaches the tunnel limit with singularities at $E=\Delta$ and above the minigap. In this limit, no secondary gap exists.}
\end{figure}

This equation can be solved analytically, however, the expression for the general solution is quite long and will not be given here. In Fig.~\ref{fig:3}, it is plotted for various values of $T$. Figure ~\ref{fig:3} (a) shows the width of the secondary gap approaching $0$ with decreasing $T$; Figure~\ref{fig:3} (b) shows the structure of the density of states above the upper edge of the minigap. Note the different scales of the energy axes in the two plots. For decreasing $T$, the LDOS approaches the tunnel limit without secondary gap but with the usual singularities \cite{golubov:89}  at $E=\Delta$ and above the minigap.

Considering $\delta=0$ provides an analytical expression for the critical phase
\begin{displaymath}
\varphi_c=\sqrt{2\left( 5 \sqrt{5}-11 \right)}T\frac{\Delta}{E_{\textrm{Th}}}.
\end{displaymath}
Similarly for $\varphi=0$ we find an analytical expression for the critical energy $\delta_c$ in the limit of large $E_{\textrm{Th}}$ describing the width of the gap 
\begin{displaymath}
\delta_c = f(T) \left(\frac{\Delta}{E_{\textrm{Th}}} \right)^2,
\end{displaymath}
$f(T)$ being a lengthy expression related to the solution of a quartic equation. For $T\ll1$ it has the form $f(T) \approx 1/2 (T/4)^4$. The position of the minigap edge (Fig.~\ref{fig:3} (b)) can as well be calculated analytically. For $T\ll1$ it is given by 
\begin{displaymath}
\delta_{mini} = (8+12 T^{2/3}) \left(\frac{\Delta}{E_{\textrm{Th}}} \right)^2.
\end{displaymath}

\subsection {Combination of transmission eigenvalues}

A generalization of the calculations from the previous section can be achieved by considering not only one constant transmission eigenvalue, but a whole set of different transmission eigenvalues, each weighted with a specific weight $w_n$. We stick to a symmetric system with only one set of transmission eigenvalues and weights $\lbrace T_n, w_n \rbrace$ describing both sides. From the huge variety of possible sets, which could be analyzed, we pick only one in order to demonstrate that the secondary gaps structure in principle is not limited to only a single gap: An even finer subdivision of the LDOS below $\Delta$ can be observed for certain contact types. We calculate the LDOS for one representative set $\lbrace T_n, w_n \rbrace$ given by

\hspace{3mm}
\begin{center}
  \begin{tabular}{ l || l | l | l | l | l | l | l | l | l | l | l | l }
    \hline
     $T_n$ & 0.1 & 0.2 & 0.3 & 0.4 & 0.5 & 0.6 & 0.7 & 0.8 & 0.9 & 1 \\ \hline
     $w_n$ & 10 & 5 & 5 & 2 & 3 & 20 & 50 & 70 & 90 & 200 \\ 
    \hline
  \end{tabular}
\end{center}
\hspace{3mm}

\begin{figure}[t] 
\vspace{5mm}
\includegraphics[width=0.95\columnwidth,angle=0]{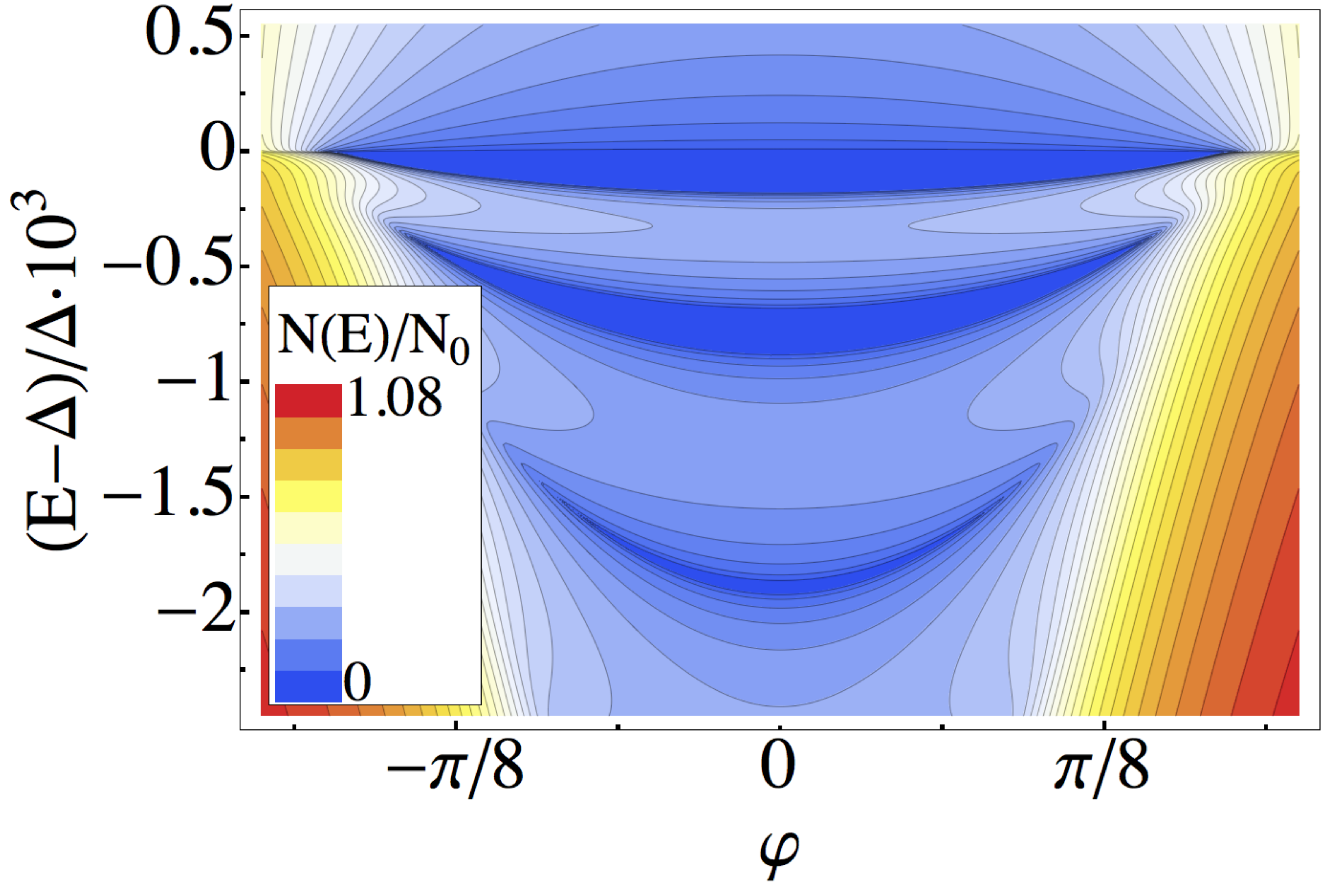}
\caption{\label{fig:4} DOS of a symmetric system described by a discrete set of transmission eigenvalues $T_n$ and weights $w_n$. In the energy range below $\Delta$ we find a multiply gapped density of states consisting of three secondary gaps with a finite DOS between them.}
\end{figure}

A plot of the numerical results in the energy range below $\Delta$ is shown in Fig.~\ref{fig:4}. We find that for certain sets of transmission eigenvalues not only one secondary gap appears, but the DOS acquires an even finer structure with multiple subgaps. The number of subgaps depends strongly on the set $\lbrace T_n, w_n \rbrace$ under consideration and on $E_{\textrm{Th}}$. In the presented case we find three gaps at $E_{\textrm{Th}}=\Delta$. Similar to the previously found secondary gaps they are symmetric around $\varphi=0$ and vanish at some critical phase, which is not the same for different subgaps.

\subsection{Continuous transmission distributions $\rho(T)$}

So far, we investigated systems where scattering in the contacts is described by constant transmission eigenvalues. However, in systems experimentally accessible scattering is rather described by a continuous transmission distribution $\rho(T)$ than by discrete transmission eigenvalues $T$. It is thus of crucial interest whether the secondary gap appears as well if continuous transmission distributions are considered. Here, we investigate three generic contact types, each characterized by a distribution of transmission eigenvalues of the form $\rho(T)\sim1/(T^\beta \sqrt{1-T})$ with $\beta=1/2, 1, 3/2$. The normalization constant is determined by the condition $G=G_Q \int_0^1 T \rho(T) dT$. The distributions for $x=3/2$ and $x=1$ correspond to a dirty and a diffusive connector \cite{schep:97,dorokhov:82} , respectively. The distribution for $x=1/2$ is equivalent to two ballistic connectors with equal conductances in series \cite{qt}.  

Again we stick to symmetric setups with equal contacts on both sides. The characteristic functions corresponding to the considered distributions have a relatively compact form \cite{belzig:00}. For a dirty connector ($x=3/2$) it is given by $X=\sqrt{(1+a)/2}$. In case of a diffusive contact ($x=1$) the characteristic function is $X=\sqrt{(1-a^2)}/\arccos{a}$. And for the double ballistic contact \cite{qt} we have $X=((1+a)/2+\sqrt{(1+a)/2})/2$.

\begin{figure}[t] 
\vspace{5mm}
\begin{overpic}[width=0.8\columnwidth,angle=0]{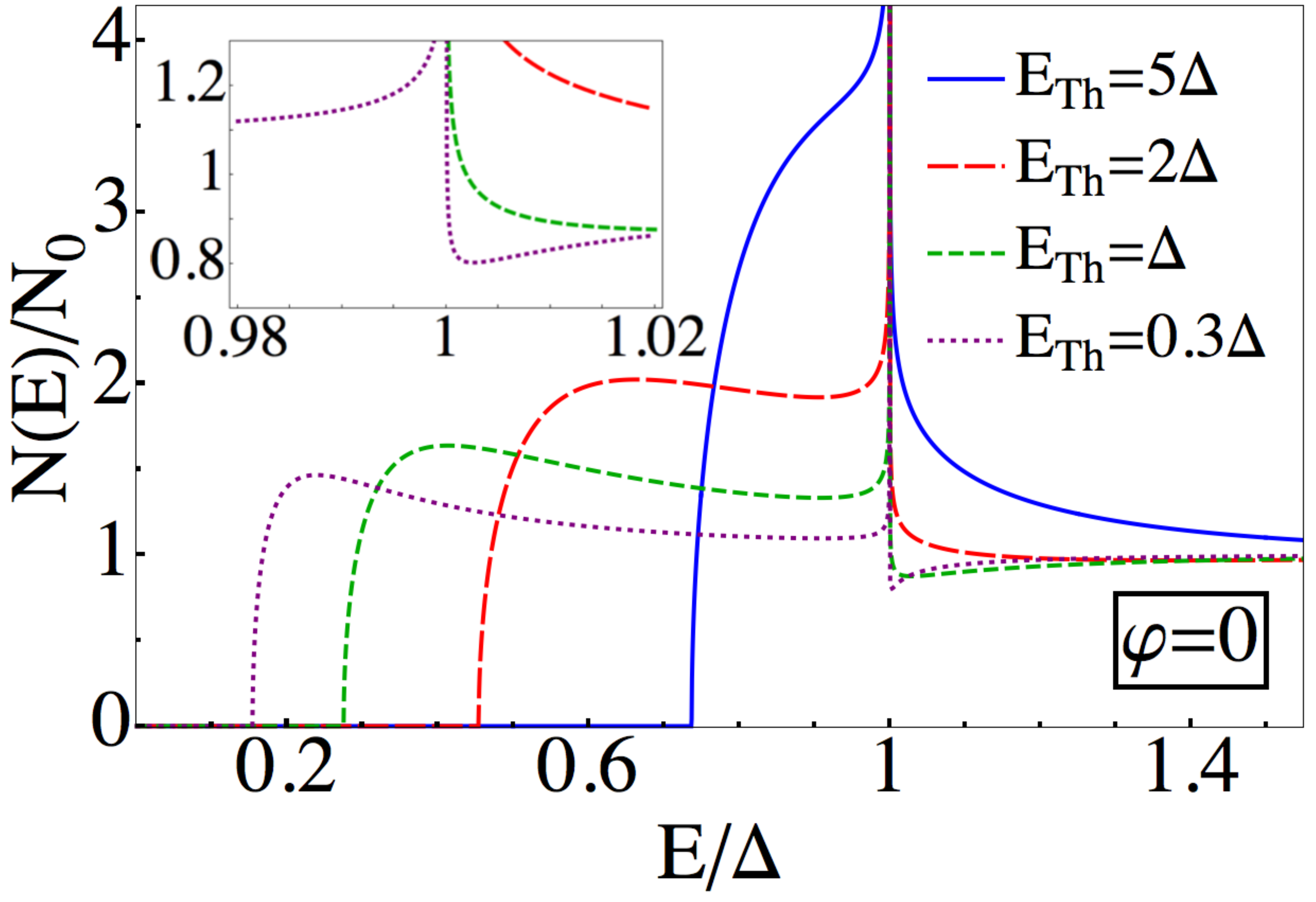}
\put(3,66){\makebox(0,3){$(a)$}}
\end{overpic}
\begin{overpic}[width=0.8\columnwidth,angle=0]{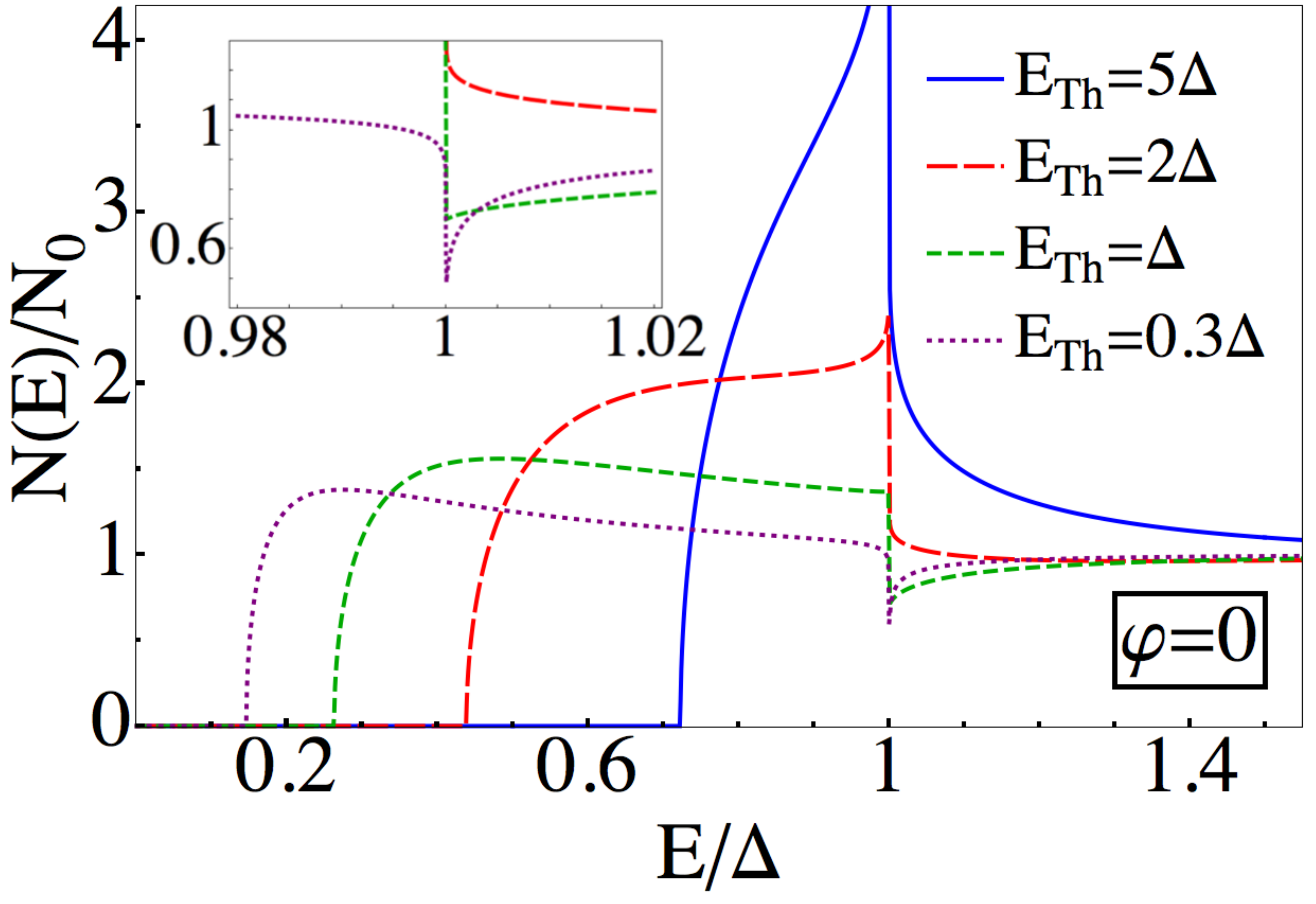}
\put(3,66){\makebox(0,3){$(b)$}}
\end{overpic}
\begin{overpic}[width=0.8\columnwidth,angle=0]{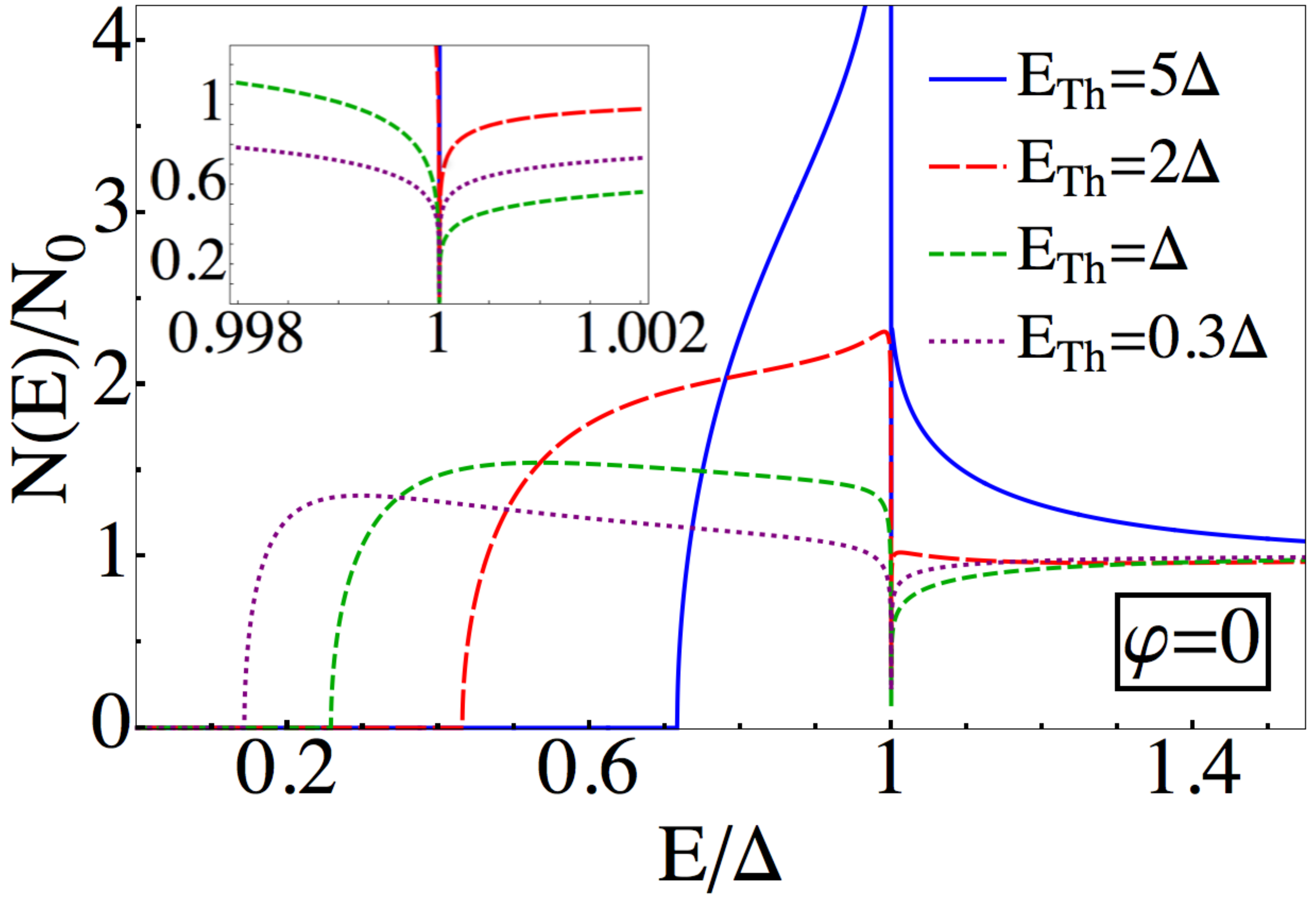}
\put(3,66){\makebox(0,3){$(c)$}}
\end{overpic}
\caption{\label{fig:5} LDOS for $\varphi=0$ for symmetric setups with contacts having continuous transmission distributions given by $\rho(T)=1/(T^x \sqrt{1-T})$ with $x=1/2, 1, 3/2$, respectively. With increasing relative weight at transmissions close to $T=1$ the LDOS at $E=\Delta$ is more and more suppressed. (a) The plots correspond to $x=3/2$ (dirty contact). The relative weight of transmission eigenvalues around 0 is the strongest of all distributions considered in this section. No suppression of the LDOS at $E=\Delta$ is found. (b) In the diffusive case ($x=1$) there is no gap either. However, the LDOS is weakly suppressed at $E=\Delta$ for $E_{\textrm{Th}} \sim 0.3 \Delta$.  (c) For double ballistic contacts, a dip in the LDOS appears, which is fully suppressed at $E=\Delta$. However, no gap of finite width is found.}
\end{figure}

Figures~\ref{fig:5}(a)--(c) show the numerical results for the LDOS calculated from Eq. (\ref{eq:3}) for the three cases. Since in previous calculations the suppression of the LDOS around $\Delta$ was strongest at $\varphi=0$, only this case is presented. Figure~\ref{fig:5} (a) contains the numerical results for dirty contacts ($x=3/2$) for different values of $E_{\textrm{Th}}$. No signature for a suppression of the LDOS at the superconducting gap edge $\Delta$ is found. The plots in Fig.~\ref{fig:5} (b) are the results for diffusive contacts ($x=1$). For $E_{\textrm{Th}} \sim 0.3 \Delta$, a weak suppression at $E=\Delta$ can be seen. The inset with a higher resolution of the energy range of interest, however, shows, that this suppression is no gap. At first glance, this seems to disagree with Ref.~\cite{levchenko:08} but is possibly due to differences in the considered geometries. The lowest Fig.~\ref{fig:5} (c) shows the results for the third contact type ($x=1/2$) with the highest weight at transmission eigenvalues around $T=1$ of all three distributions. We find a strong suppression at $E=\Delta$ for all considered values of $E_{\textrm{Th}}$. The inset of this plot confirms that the LDOS is suppressed to 0 at $E=\Delta$.

In summary we find that with decreasing weight of $\rho(T)$ at $T=0$ the suppression of energy levels at $E=\Delta$ is reinforced. This observation supports the idea that the existence of Andreev bound states with energies directly below $E=\Delta$ is related to the tunnel character of the boundaries, i.e., the transport channels with transmissions at $T\sim 0$.

\begin{figure}[t] 
\vspace{5mm}
\includegraphics[width=0.8\columnwidth,angle=0]{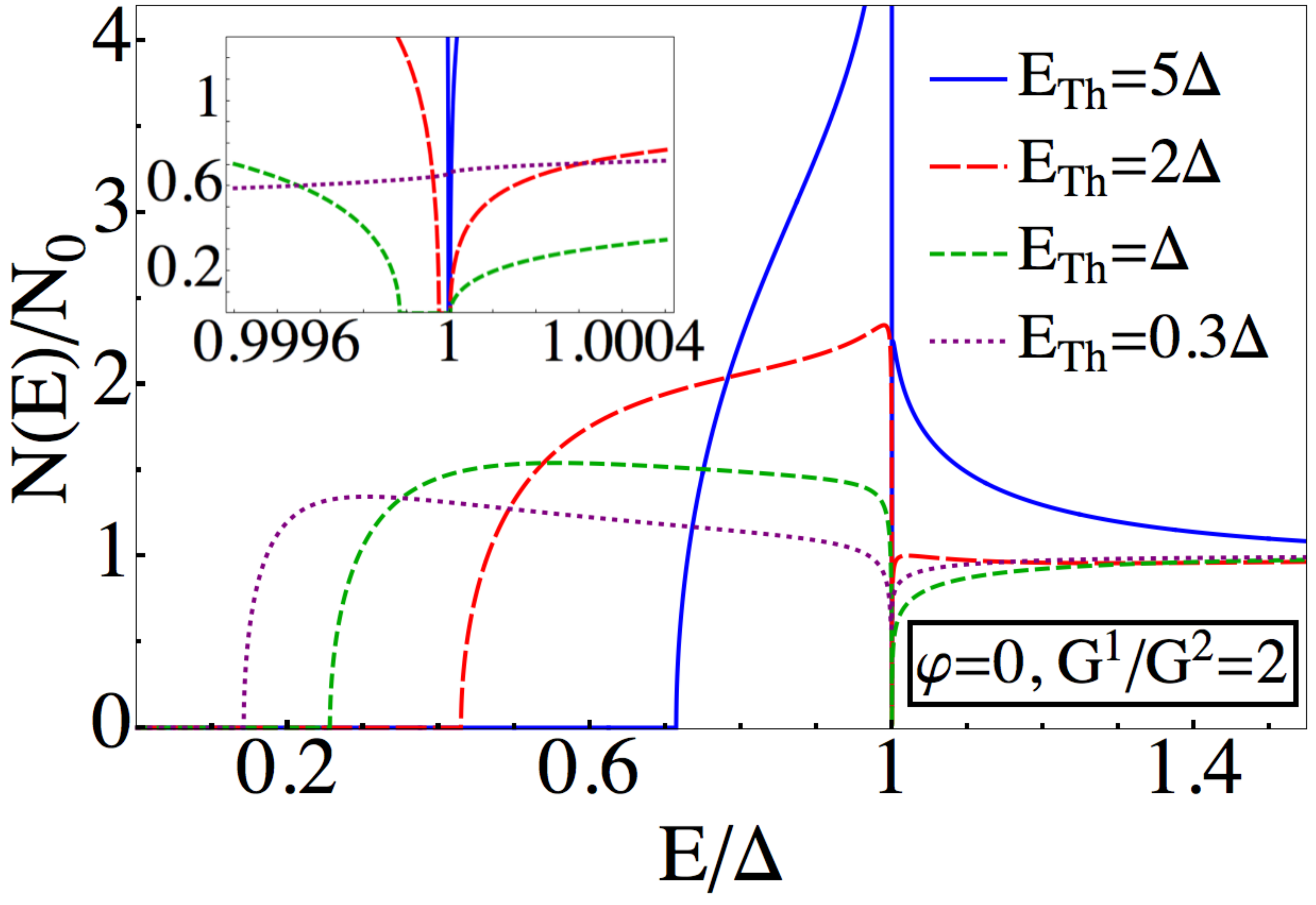}
\caption{\label{fig:6} LDOS at $\varphi=0$ for a relation of conductances $G^1/G^2=2$ for different values of $E_{\textrm{Th}}$. A $E_{\textrm{Th}}$-dependent secondary gap is found below $E=\Delta$.}
\end{figure}

In order to check this, a fourth type of transmission distribution is considered. Each contact in this system is built up of two ballistic contacts in series ($G_i^1$ and $G_i^2$) having different conductances. Note that the total setup is still symmetric. The corresponding transmission distribution fundamentally differs from the previously considered distributions in the sense that it has no contribution at small transmissions, i.e., below a critical value $T_{min}$ given by $T_{min}=(G^1_i-G^2_i)^2/(G^1_i+G^2_i)^2$ . For $G^1_i=G^2_i$ it follows $T_{min}=0$ which corresponds to the previously investigated distribution with $x=1/2$. Since we again stick to symmetric setups we drop the lead index $i$ in the following. The characteristic function is given by \cite{qt}
\begin{displaymath}
X(a)=\frac{G^1 G^2}{(G^1+G^2)^2}\frac{a-1}{1-\sqrt{1-\frac{4G^1 G^2}{(G^1+G^2)^2} \frac{a-1}{a+1}}}.
\end{displaymath}
The results are plotted in Fig.~\ref{fig:6}. Compared to the previous  distributions, a gap of finite width appears directly below $E=\Delta$ (inset of Fig.~\ref{fig:6}). As expected, the gap appears when there is no contribution of the transmission distribution around $T=0$. This agrees with the results of Sec. III. A. In the following we derive a criterion that relates the existence of the secondary gap directly below $E=\Delta$ to the weight of the transmission distribution around $T=0$.

\subsection{Analytical criterion}

In this section, we show that the existence of the secondary gap below $\Delta$ is indeed directly related to the distribution $\rho(T)$ of the transmission eigenvalues in the vicinity of $T=0$. We show that the existence of a minimal transmission $T_{min} > 0$ results in a secondary gap below $\Delta$. To reach this conclusion, we consider Eq. (\ref{eq:3}) for a symmetric setup containing one general function $X(a)$ and $\varphi=0$. It reads as
\begin{displaymath}
\frac{g}{f}=\frac{c(E)}{s(E)}-\frac{2iE}{E_{\textrm{Th}} s(E)}X(a).
\end{displaymath}
We linearize this equation in $\delta=(\Delta-E)/\Delta$ in the limit $E_{\textrm{Th}} \gg \Delta$. The left side is expanded in $1/g$ which must be small in order to be valid. This has to be verified for the solution. Using the limiting form of $c(E)$ and $s(E)$ at small $\delta$ we get
\begin{displaymath}
\frac{1}{2 g^2}=-\delta+\frac{2\Delta}{E_{\textrm{Th}}}\sqrt{2\delta}X(a).
\end{displaymath}
In the leading order in $\delta$, the general expression for $a(f, g, E)$ given after Eq. (5) yields:  $a=i/\sqrt{2\delta} \left (\delta g -1/(2g) \right )$. Introducing rescaled variables we get an equation without explicit dependence on $E_{\textrm{Th}}$. With the definitions $\delta=x (\Delta/E_{\textrm{Th}})^2$ and $g=y(\Delta/E_{\textrm{Th}})^{-1}$ we have
\begin{equation}
\label{eq:analytic1}
\frac{1}{2 y^2}=-x+2\sqrt{2x}X\left (\frac {i}{\sqrt{2x}} \left (x y-\frac{1}{2 y} \right )\right).
\end{equation}
The LDOS is related to the real part of $g$, so a purely imaginary solution for $y(x)$ at small $x$ means having a gap at energies E close to $\Delta$. To show this we consider the sum $D(a)$ in Eq. (\ref{eq:characteristic_funct}), which is an integral in the continuous case. For an arbitrary distribution $\rho(T)$, which can be normalized to satisfy $\int \rho(T) T dT=1$, it reads
\begin{displaymath}
D(a)=\int_{T_{\textrm{min}}}^1 \frac{\rho(T) dT}{\frac{1}{T}+\frac{1}{2}\left(\frac {i}{\sqrt{2x}} \left (x y-\frac{1}{2 y} \right )-1\right)} ,
\end{displaymath}
where the minimal transmission in $\rho(T)$ was used to replace the lower boundary in the integral. Considering small $x$ and assuming $y\sim x^\alpha$ with $\alpha>-1/2$ (later verified by the solution) we can neglect all other terms in the denominator compared to $-i/(4\sqrt{2x}y)$. We find
\begin{displaymath}
X(a)= {-ik}/{4 \sqrt{2x}y},
\end{displaymath}
$k$ being a constant factor defined as $k=1/{\int \rho(T) dT}$. It is constrained to the interval $\rbrack 0, 1 \rbrack$, $k=1$ corresponding to a ballistic setup. Its value depends on the exact form of $\rho(T)$ in the whole interval, particularly on $T_{min}$. For a diffusive connector which is cut at $T_{min}$ we have $k=2 \sqrt{1-T_{min}}/[2\ln(1+\sqrt{1-T_{min}})-\ln{T_{min}}]$, which becomes $1$ for $T_{min}=1$ and approaches $0$ as $T_{min}$ approaches $0$. With this Eq. (\ref{eq:analytic1}) reduces to a quadratic equation
\begin{displaymath}
\frac{1}{2 y^2}=-x-\frac{i}{2y}k
\end{displaymath}
with solutions $y_{\pm}=(i/2)(-k/2x \pm \sqrt{k^2/4x^2+2/x})$. Both solutions are purely imaginary for $x>0$ signifying a gap in the LDOS. However only the solution $y_+ \approx i/k$ for $x \ll k$ is consistent with the previously made assumption $y(x)\sim x^\alpha$ with $\alpha>-1/2$ for small $x$. Since this solution is finite for small $x$ the second assumption, which assumed $g$ to be large, can always be fulfilled for $k>0$ by choosing $E_{\textrm{Th}}$ sufficiently large. This is in agreement with our numerical results which predict no secondary gap below some critical value of $E_{\textrm{Th}}$. 
This critical $E_{\textrm{Th}}$ depends on the value of $k$ and thus on the whole transmission distribution $\rho(T)$.
The condition $k>0$ is related to the existence of a $T_{\textrm{min}}>0$, since only $\rho(T\ll 1)\sim T^{-\alpha}$ with $\alpha>1$ leads to $k=0$. To conclude this section, we have shown that for an arbitrary transmission distribution $\rho(T)$ without contribution in a finite interval above $T=0$ a secondary gap appears directly below $E=\Delta$, if $E_{\textrm{Th}}$ is sufficiently large. 

\subsection{Asymmetric setup}

In this section, we demonstrate that having the distribution function of transmission coefficients $\rho (T)=0$ at $T$ below some $T_{min}$ is a sufficient but not necessary condition for having a secondary gap. We consider a device with two nonidentical junctions and show that the secondary gap may exist even if the latter condition on $\rho(T)$ is violated. The secondary gap, however, does not appear directly below $\Delta$ but is shifted to slightly smaller energies, thus it appears in a different regime than considered in Sec. III D.

\begin{figure}[t] 
\vspace{10mm}
\begin{overpic}[width=0.95\columnwidth,angle=0]{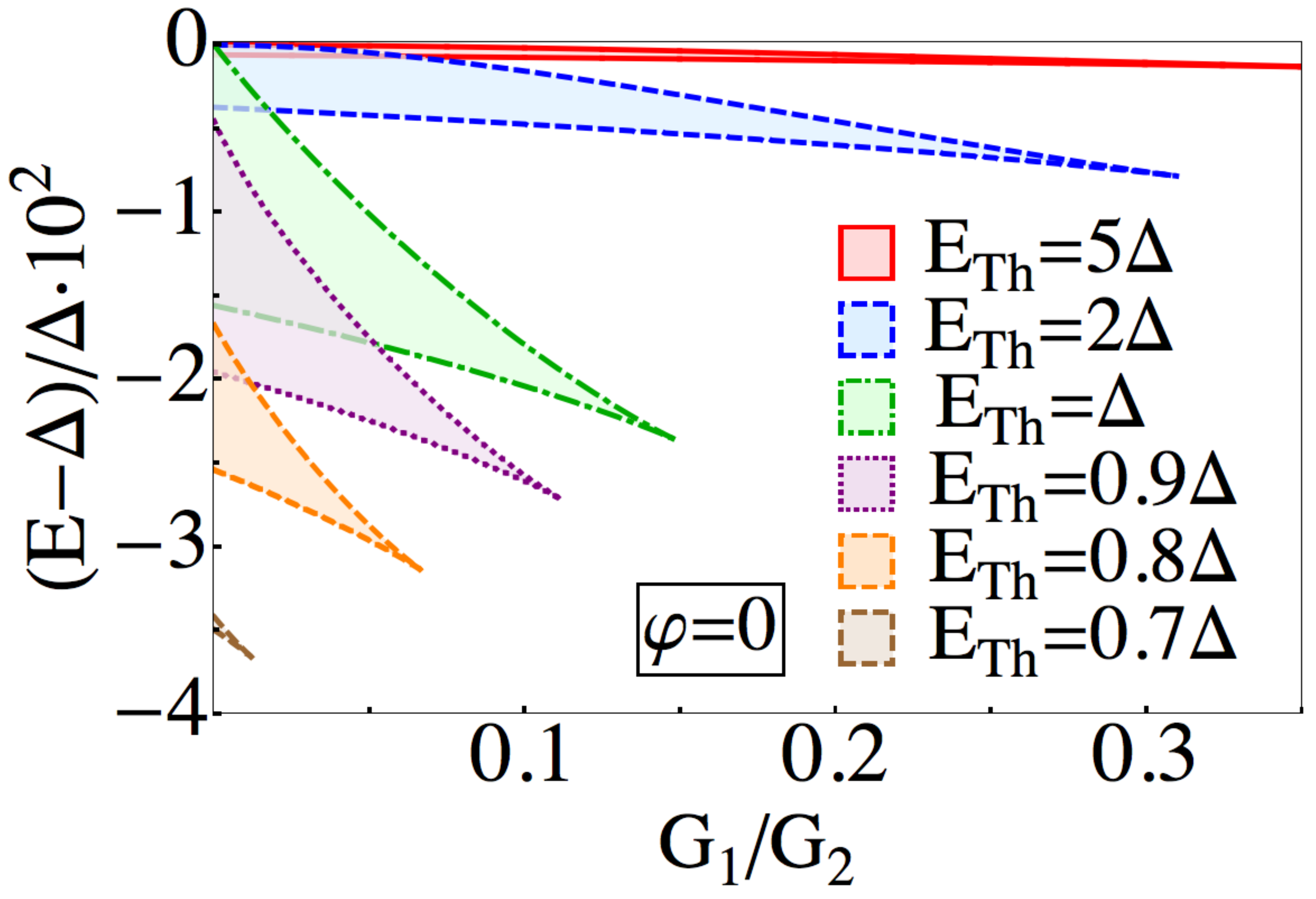}
\put(3,66){\makebox(0,3){$(a)$}}
\end{overpic}
\begin{overpic}[width=0.95\columnwidth,angle=0]{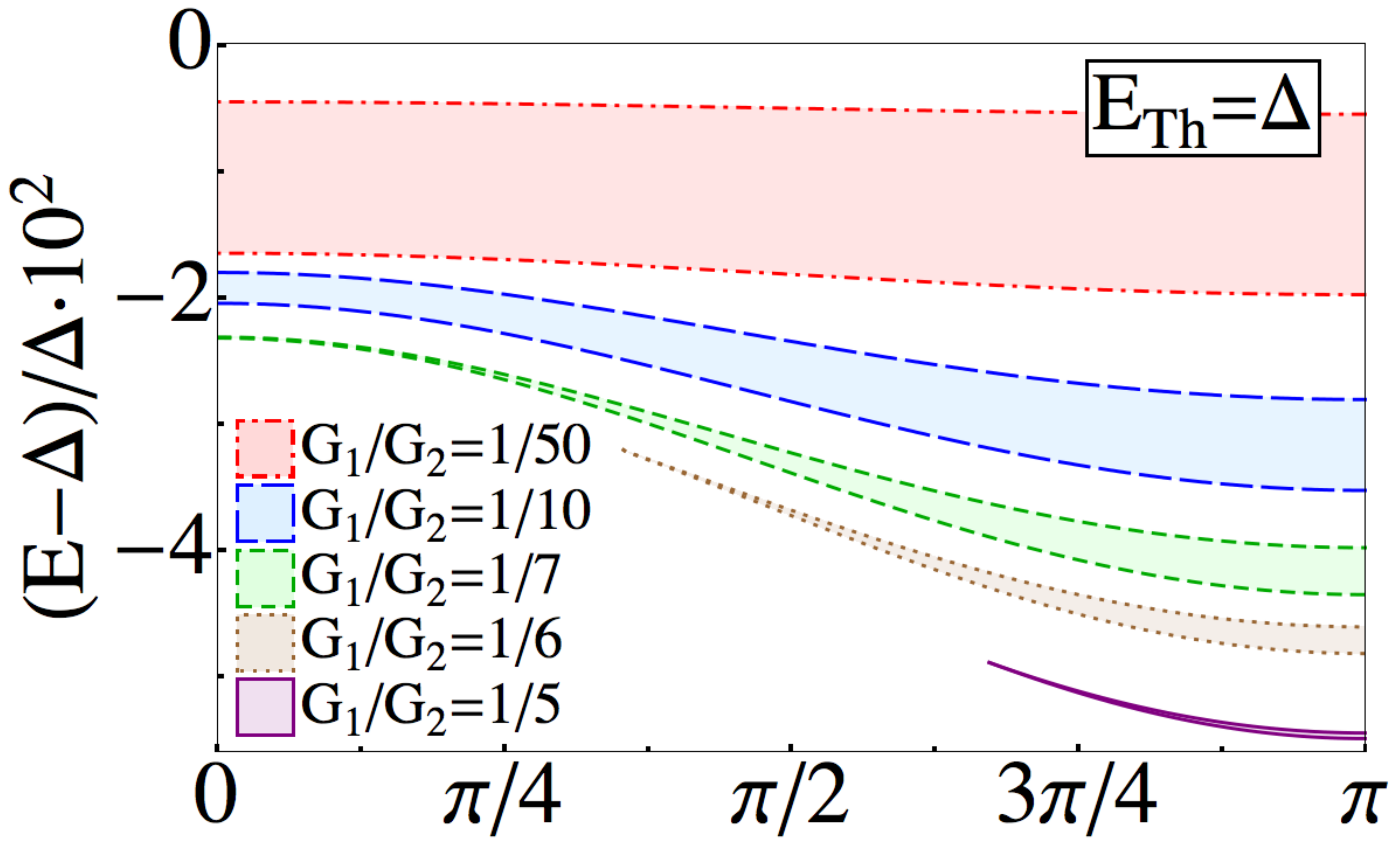}
\put(3,66){\makebox(0,3){$(b)$}}
\end{overpic}
\caption{\label{fig:7} Asymmetric setup with a tunnel contact ($G_1$) and a ballistic contact ($G_2$). (a) In the ballistic limit $G_1/G_2 \ll 1$ a secondary gap exists below $\Delta$, which vanishes as $G_1/G_2$ increases. The shaded regions denote the gap. (b) Phase dependence of the secondary gap for constant $E_{\textrm{Th}}=\Delta$ and different values of $G_1/G_2$.}
\end{figure}

The most interesting case to which we confine ourselves here is the one which combines the two extremal contact types: A tunnel contact with $T \sim 0$ for all transport channels on one side and a ballistic contacts with $T=1$ for all channels on the other side. The conductances of both sides enter our calculation via the relation $G_1/G_2$, where $G_1$ denotes the conductance of the tunnel contact and $G_2$ corresponds to the ballistic contact. For $G_2 \gg G_1$ the role of the tunnel contact is negligible and the result of a symmetric ballistic system at $\varphi=0$ showing a secondary gap below $\Delta$ \cite{reutlinger:14} should be reproduced. It is of particular interest how the transition from the ballistic limit $G_1/G_2 \ll 1$ to the tunnel limit $G_1/G_2 \gg 1$ occurs. Results of the previous section indicate that there should be no secondary gap just below $\Delta$: The condition on $\rho(T)$ for the appearance of the secondary gap is violated by the presence of a tunnel junction even for $G_1/G_2 \ll 1$.

Figure~\ref{fig:7} shows the results of our numerical calculation. In Fig.~\ref{fig:7} (a), the phase $\varphi$ is fixed at $\varphi=0$ and the vanishing of the secondary gap is shown for different values of $E_{\textrm{Th}}$ as $G_1/G_2$ increases. At small $G_1/G_2$, the upper gap edge is close to its ballistic value and decreases as $G_1/G_2$ increases. Similarly, the lower gap edge is close but slightly below its value of the symmetric ballistic case and decreases with increasing $G_1/G_2$. At some critical value of $G_1/G_2$ which depends on $E_{\textrm{Th}}$ the secondary gap disappears. This critical value is smaller for smaller $E_{\textrm{Th}}$. In Fig.~\ref{fig:7}(b), the Thouless energy is fixed at $E_{\textrm{Th}}=\Delta$ and the phase dependence of the secondary gap is plotted for different values of $G_1/G_2$. We find that for the considered system the secondary gap has its maximum width not at $\varphi=0$ as one might expect, but at $\varphi=\pi$. With decreasing $G_1/G_2$, the $\varphi=0$ result of the symmetric ballistic system is approached at all phases. Compared to previous findings for asymmetric ballistic contacts \cite{reutlinger:14}, where a similar behavior with decreasing $G_1/G_2$ was found, no band of finite DOS separates the gap at $\varphi=0$ from the gap at $\varphi=\pi$.

\subsection{Spatial dependence}

\begin{figure}[h]
\begin{center}
\includegraphics[width=0.99 \columnwidth]{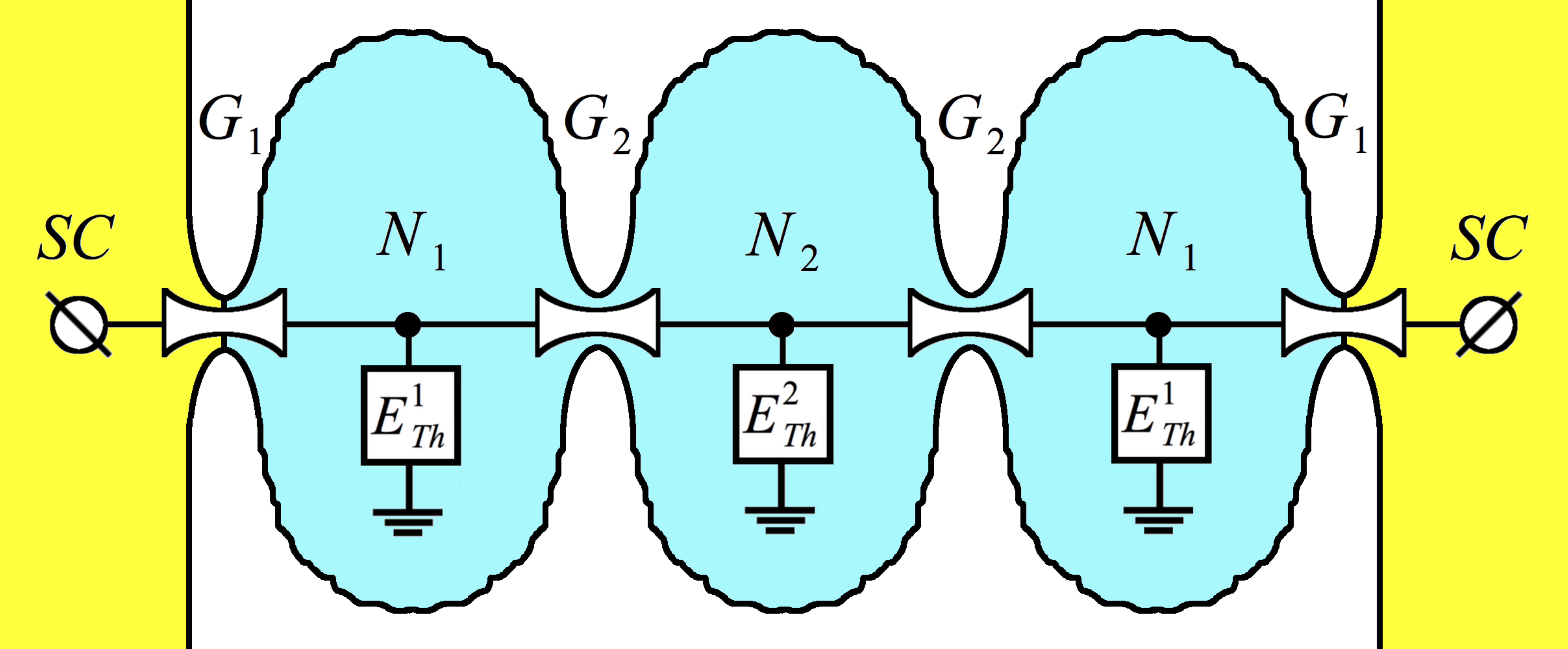}
\caption{\label{fig:8} Sketch of the system with three normal nodes connected to two superconductors at $\varphi=0$. Such a geometry can model, for example, a series of three cavities connected by point contacts of different widths. The contacts have the conductances $G_1$ and $G_2$ and at each normal node a leakage current described by $E_{Th}^{1,2}$ is taken into account.}
\end{center}
\end{figure}

In order to achieve a spatial resolution of the local density of states (LDOS) we consider a symmetric model of three normal islands connected to two superconductors at $\varphi=0$. Due to symmetry the Green's functions in the left and right normal nodes are equal. Both nodes are thus called $N_1$ in the following, the central node is called $N_2$. The nodes $N_1$ are connected via a ballistic conductance $G_1$ to the superconductor and via $G_2$ to $N_2$. In each normal node, electron-hole-decoherence is described through Thouless energies $E_{Th}^{1,2}$, respectively. The system setup is sketched in Fig.~\ref{fig:8}.

\begin{figure}[t] 
\vspace{5mm}
\begin{overpic}[width=0.9\columnwidth,angle=0]{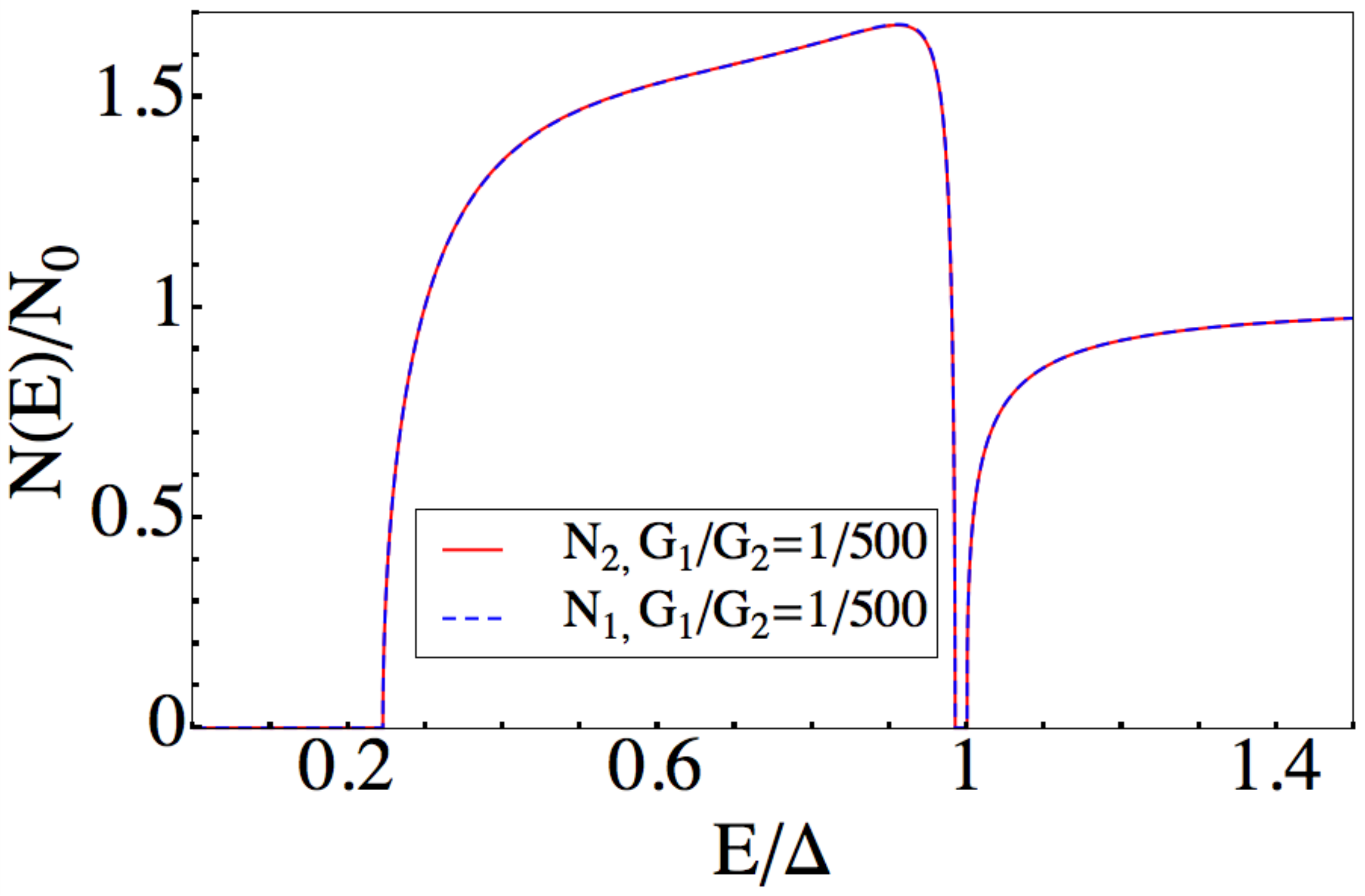}
\put(3,66){\makebox(0,3){$(a)$}}
\end{overpic}
\begin{overpic}[width=0.9\columnwidth,angle=0]{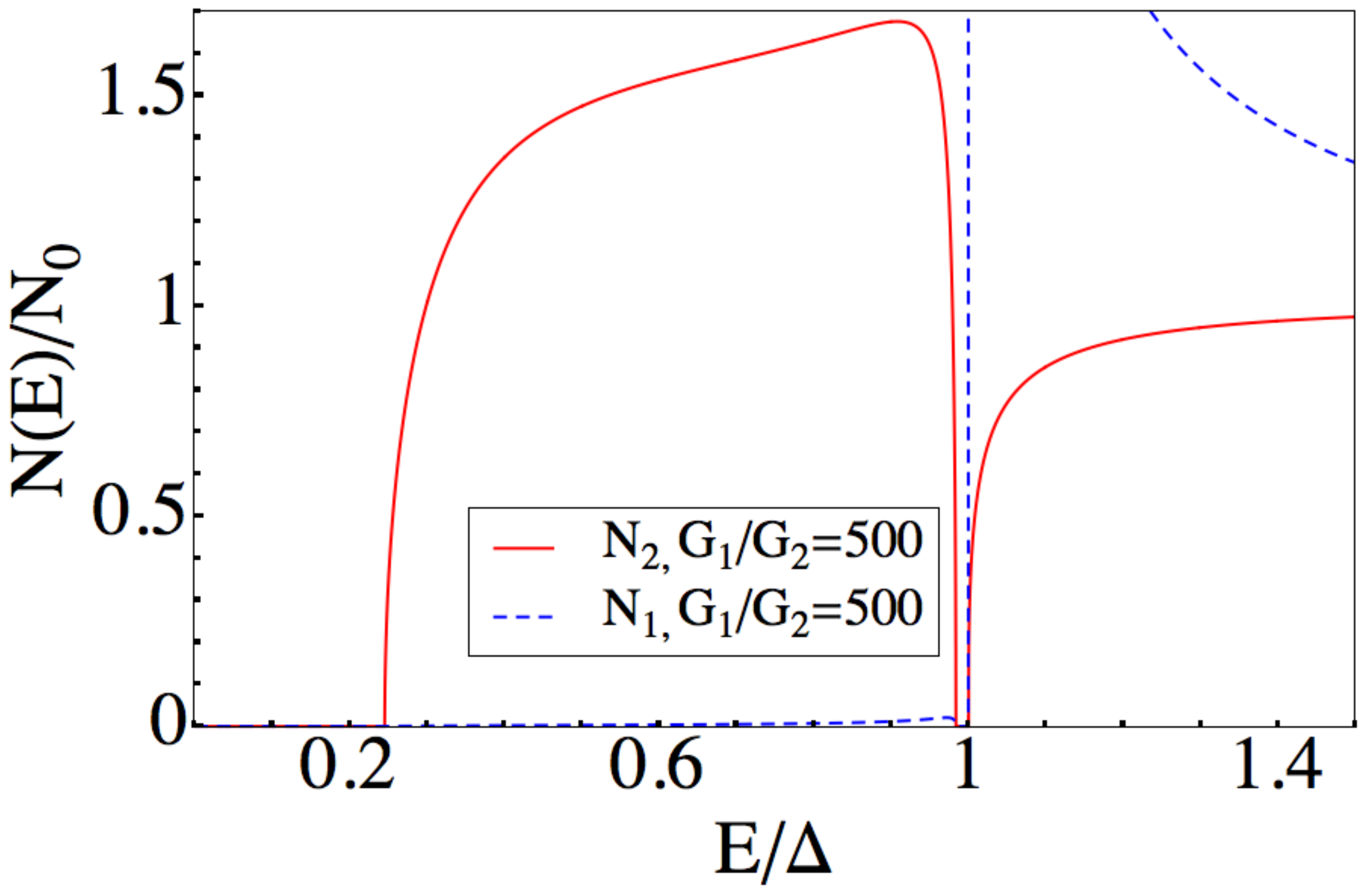}
\put(3,66){\makebox(0,3){$(b)$}}
\end{overpic}
\caption{\label{fig:9} Spatial dependence of the DOS. (a) The DOS for $G_1/G_2=1/500$ is constant in the normal region. (b) For $G_1/G_2=500$ the outer nodes are strongly coupled to the superconductors, whereas the inner node shows the standard result for a ballistic cavity (a).}
\end{figure}

Matrix current conservation in one of the nodes $N_1$ and in the node $N_2$ determines the Green's function and the LDOS in the particular node \cite{nazarov:94, nazarov:99}:
\begin{align*}
	\label{eq:curr_cons}
 	\hat I_{1S}+\hat I_{12} +i(G_1+G_2)(E/E_{Th}^1)[\hat\tau_3,\hat G_1(E)]&=0,\\
 	-2 \hat I_{12} +i(2 G_2)(E/E_{Th}^2)[\hat\tau_3,\hat G_2(E)]&=0.
\end{align*}
The matrix currents $\hat I_{12}$ and $\hat I_{1S}$ are defined according to the definition (\ref{eq:curr}). In general, the system is described by three parameters $E_{Th}^1$, $E_{Th}^2$, and $G_1/G_2$. We fix one parameter by considering only systems with equal mean level spacings in all three normal nodes $\delta_s^1=\delta_s^2=\delta_s$. Furthermore, we fix $G_1 G_2/((G_1+3G_2)\Delta) \delta_s=G_Q/2$. In this case for $G_1/G_2 \ll 1$ the LDOS in $N_1$ and $N_2$ are equal and correspond to the result of Figure 1(a) in \, \cite{reutlinger:14}, and for $G_1/G_2 \gg 1$ we have the BCS-DOS in $N_1$ and again the result of Figure 1(a) of \, \cite{reutlinger:14} in $N_2$. Figure~\ref{fig:9} shows the LDOS in the two nodes for $G_1/G_2=1/500$ [Fig.~\ref{fig:9} (a)] and for $G_1/G_2=500$ [Fig.~\ref{fig:9} (b)]. Taking this into account, $E_{Th}^1$ and $E_{Th}^2$ can be expressed in terms of $G_1/G_2$. We find
\begin{align*}
E_{Th}^1 &= (G_1+G_2)/G_Q \delta_s=\Delta (G_2+G_1) \\
&\times (3 G_2+G_1)/(2 G_1 G_2),\\
E_{Th}^2 &= 2 G_2/G_Q \delta_s=\Delta (3 G_2+G_1)/G_1.
\end{align*}

\begin{figure}[t]
\vspace{5mm}
\begin{overpic}[width=0.9\columnwidth,angle=0]{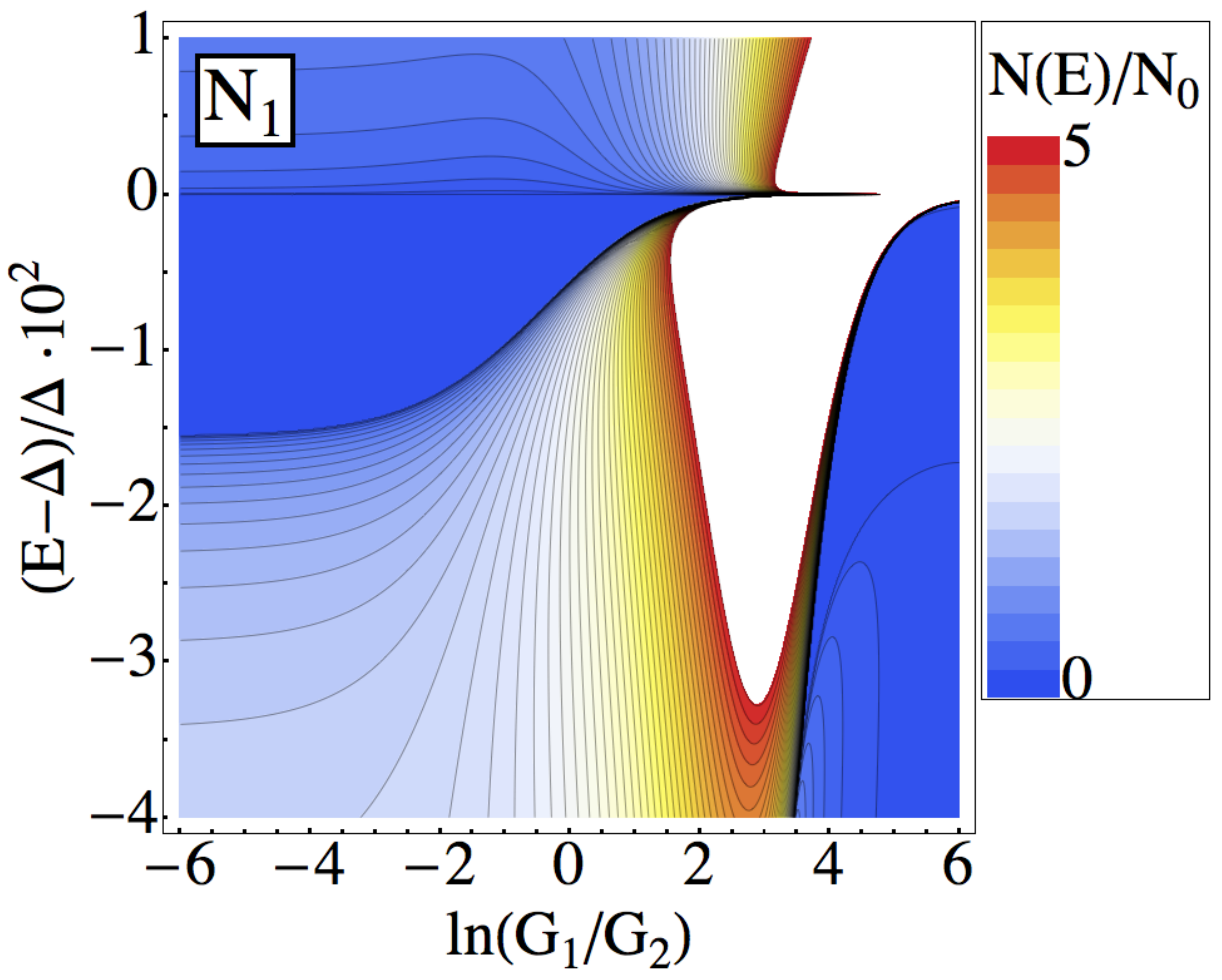}
\put(3,80){\makebox(0,3){$(a)$}}
\end{overpic}
\begin{overpic}[width=0.9\columnwidth,angle=0]{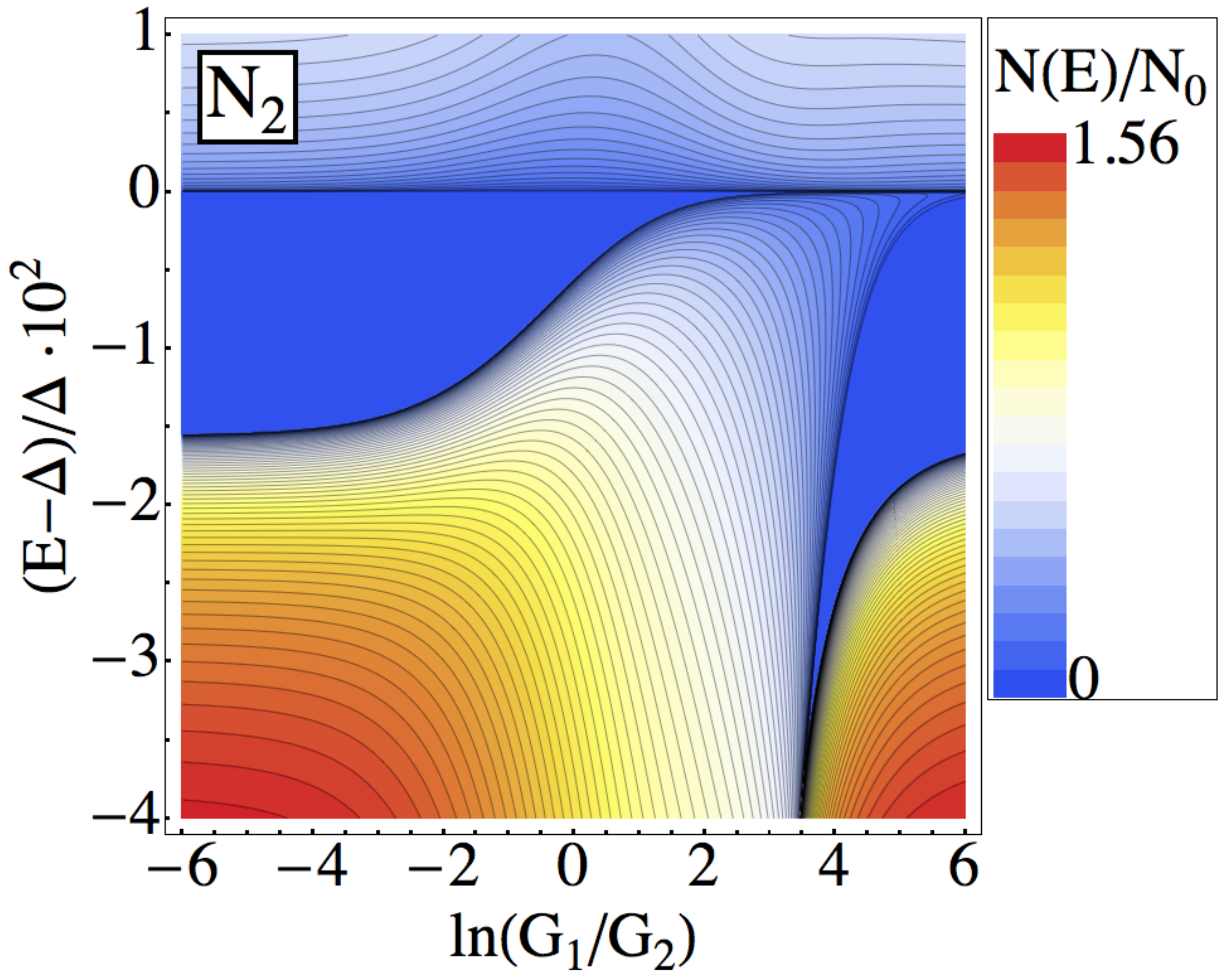}
\put(3,80){\makebox(0,3){$(b)$}}
\end{overpic}
\caption{\label{fig:10} LDOS in $N_1$ (a) and $N_2$ (b) for intermediate values of $G_1/G_2$ between $e^{-6}$ and $e^6$. The white regions in (a) are regions where $N(E)/N_0>5$. Whenever a gap appears in the central node $N_2$ there appears a gap in the outer nodes $N_1$ as well.}
\end{figure}

In Fig.~\ref{fig:10}, we show the numerical results for the LDOS in both normal nodes in dependence of $G_1/G_2$ and energy in the secondary gap region below the superconducting gap edge $\Delta$. Figure~\ref{fig:10} (a) shows the result in the outer nodes $N_1$, and Fig.~\ref{fig:10} (b) shows the result for the inner node $N_2$. The white region in Fig.~\ref{fig:10} (a) denotes $N(E)/N_0>5$. The main finding of our calculations is that the LDOS in the two nodes differs only where $N(E)/N_0>0$ in both nodes. Whenever it is zero in the central node it is also zero in the outer nodes and vice versa. We thus find a behavior of the secondary gap similar to what is already known from the usual minigap \cite{pilgram:00}. The width of this gap is not position dependent, only the LDOS above/below the particular gap edge varies with position. 

We thus expect the secondary gap not only in the LDOS of a singular point, but as well in the integrated DOS of a finite region. Depending on the parameters, not for every system does a secondary gap appear. However, if it appears in one point, it exists also in every other point of the normal part. The previously used model with only a single normal node between the superconductors is thus sufficient if the main interest concerns the existence of the secondary gap and its properties. However, with this method we cannot calculate a position-resolved LDOS and thus cannot make statements about the integrated DOS in the energy interval between minigap and secondary gap.

\section{1D Scattering Model}

\begin{figure}[t] 
 \includegraphics[width=0.7\columnwidth,angle=0]{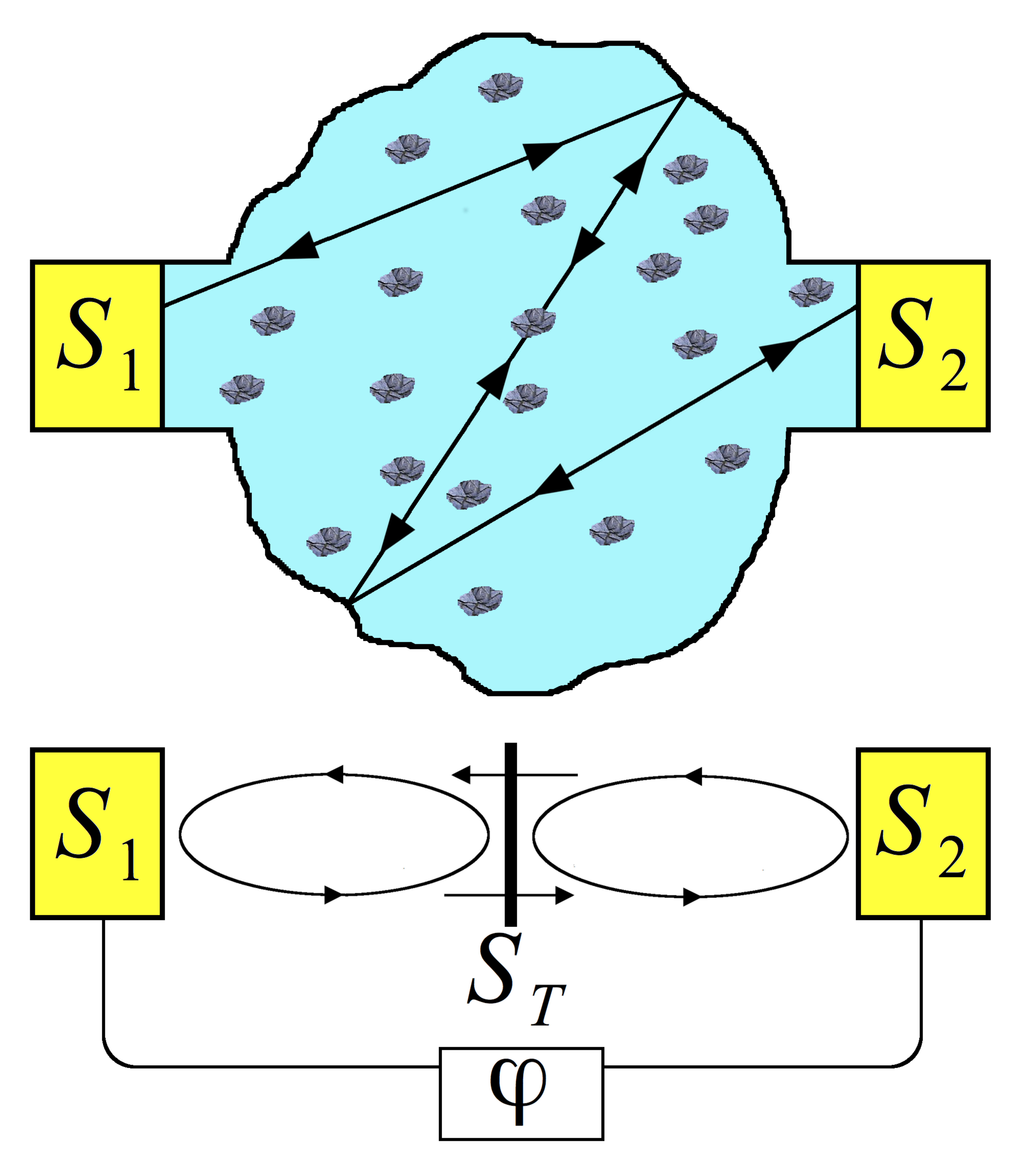}
\caption{\label{fig:11}
Upper plot: Sketch of the system with a dirty normal metal between two superconductors $S_1$ and $S_2$. The upper plot shows one possible path connecting left and right superconductors with a single scattering event. In the lower plot, the description of excitations following such paths is shown in terms of an impurity scattering matrix ($S_T$) in the normal region and Andreev reflection at the superconductors.
}
\end{figure}

The secondary gap we found for diffusive Josephson systems was calculated using Green's function techniques in the quasiclassical approximation. Whereas this method is very powerful in calculating expectation values of physical observables, it does not provide  a simple intuitive explanation for the absence of Andreev levels in the secondary gap region and the dependence of these levels on the phase difference $\varphi$ between the superconductors. In this section, we investigate a simple 1D scattering model which is able to explain qualitatively the secondary "smile"-gap. However, since we deal with diffusive or chaotic scattering systems with large conductance, we should not expect to reproduce the details of 3D solutions.

\subsection{Single-trajectory Andreev level}

We consider a semiclassical path between the left ($S_1$) and the right ($S_2$) superconductor [Fig.~\ref{fig:11}] and first recall the characteristics of Andreev bound states between superconductors on a ballistic trajectory. The bound state energies follow from the semiclassical quantization condition:
\begin{equation}
	2E/E_{\textrm{Th}}-2\arccos(E/\Delta) \pm \varphi=2\pi n\,.
	\label{eq:bohr_sommerfeld}
\end{equation}
Here, the first term is the phase difference acquired between electron and hole upon traversing the normal region. $E_{\textrm{Th}}$ is essentially the inverse traversal time, which could also be due to ballistic motion $\sim d/v_F$ for a trajectory of length $d$. The second factor is twice the energy-dependent Andreev reflection phase and the third term the phase difference $\varphi$ between the superconducting order parameters. All terms together have to add to an integer multiple of $2\pi$.

This equation reproduces two limiting cases. In long junction limit $E_\textrm{Th}\ll \Delta$, we replace $\arccos(E/\Delta)\approx \pi/2$ and find the usual spectrum of Andreev levels
$E_\textrm{n}(\varphi)=(E_\textrm{Th}\pi/2)(2n+1 \pm \varphi/\pi)$. 
In this case, levels move up and down in energy linearly with the phase difference $\varphi$. The lowest positive energy states have the energies $(E_{\textrm{Th}}\pi/2)(1 \pm \varphi/\pi)$. The levels split with $\varphi$ and cross 0 at $\varphi=\pm \pi$, corresponding to the closing of the minigap. In the opposite, short junction limit $E_\textrm{Th}\to \infty$,  we neglect the first term in Eq.~(\ref{eq:bohr_sommerfeld}) and find the Andreev levels
$E(\varphi)=\pm \Delta\cos(\varphi/2)$.

The most interesting case is the "not-so-short" junction limit $E_\textrm{Th}\gtrsim\Delta$. Assuming the energy is close to $\Delta$, we can replace $E$ by $\Delta$ in the first term of Eq.~(\ref{eq:bohr_sommerfeld}), and taking $n=0$ we obtain
\begin{align}
	E(\varphi)&=\Delta\cos(\Delta/E_{\textrm{Th}}\pm\varphi/2)\nonumber\\
	&\approx \Delta\left[1-(\Delta/E_{\textrm{Th}}\pm\varphi/2)^2/2\right]\,.
\end{align}
Thus, we obtain two states shifted in phase by the (small) parameter $\Delta/E_{\textrm{Th}}$. They touch the gap at the critical phase $\varphi_c=\pm 2\Delta/E_{\textrm{Th}}$ and the maximal distance to $\Delta$ (at $\varphi=0$) is $\Delta^3/2E_{\textrm{Th}}^2$. This is in quantitative agreement with the characteristics of the secondary gap found previously within the quasiclassical Green's function theory. Note that in the present approximation, the two levels cross at $\varphi=0$. We can expect that finite backscattering will lead to an anticrossing and the phase dependence of the level resembles the "smile" shape of the secondary gap.

\subsection{Single-trajectory Andreev level with scattering}

We investigate a simple model for the anticrossing and calculate the Andreev bound-state energies for a 1D-model with impurity scattering modeled by a scattering matrix. Although this model takes only backward scattering into the same trajectory into account and neglects the complex interference effects of three-dimensional impurity scattering which are covered by our original Green's function calculations, the results provide an understanding of the phase-dependent Andreev level density of states. The bound-state energies are obtained from the scattering matrices in the normal region \cite{qt}. We consider the geometry shown in Fig.~\ref{fig:11}. The normal scattering matrix encompasses the back scattering at the impurity as well as the dynamical phases along the trajectory to the superconductor and is given by
\begin{displaymath}
	S_N^e(E,x)=
\begin{pmatrix}
 r e^{2i x E/E_{\textrm{Th}}} & t e^{i E/E_{\textrm{Th}}}\\
 t e^{i E/E_{\textrm{Th}}} & -r e^{2i (1-x) E/E_{\textrm{Th}}}
\end{pmatrix},
\end{displaymath}
where $x\in[0,1]$ accounts for the position of the impurity along the path and $t^2=T=1-r^2$ is the transmission probability. The normal region scattering matrix for holes is related  through $S_N^h(E)=S_N^{e*}(-E)$.

\begin{figure}[t]
\vspace{5mm}
\begin{overpic}[width=0.9\columnwidth,angle=0]{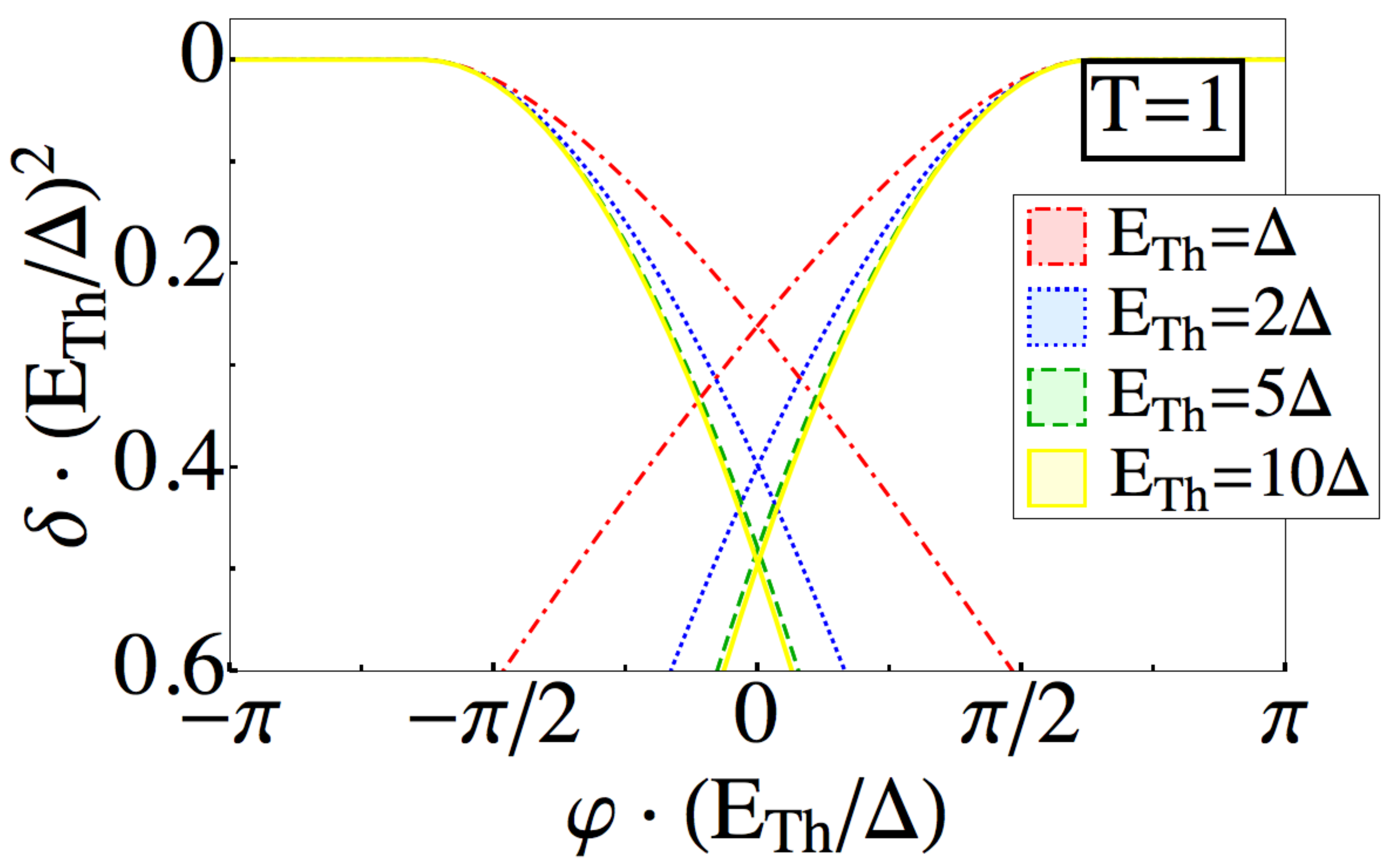}
\put(3,60){\makebox(0,3){$(a)$}}
\end{overpic}
\begin{overpic}[width=0.9\columnwidth,angle=0]{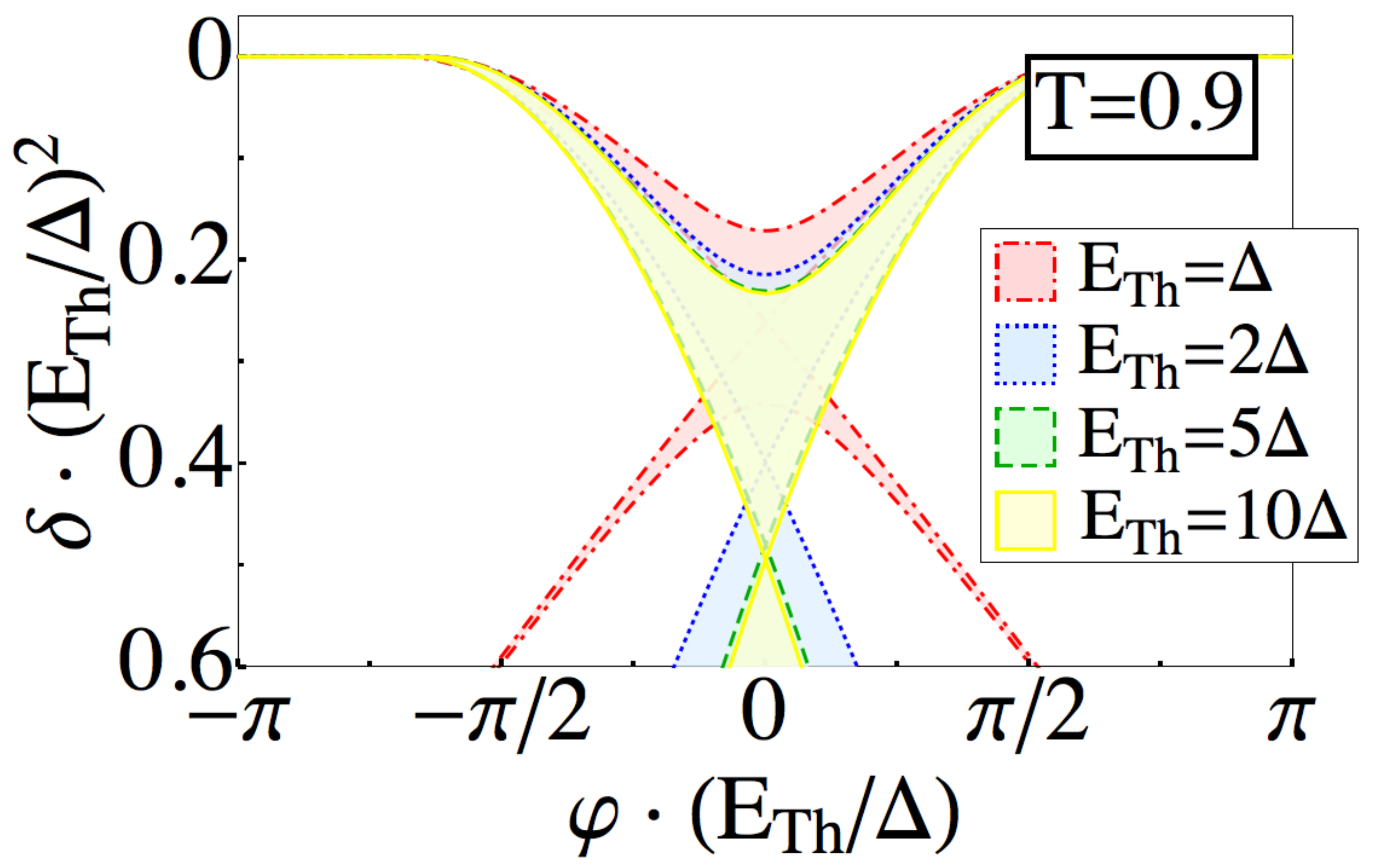}
\put(3,60){\makebox(0,3){$(b)$}}
\end{overpic}
\caption{\label{fig:12}
Energy of Andreev levels for a single mode with transmission probability $T$ [$T=1$ in (a) and $T=0.9$ in (b)] through the normal part for  $E_{\textrm{Th}}=\Delta$ (red curve), $E_{\textrm{Th}}=2\Delta$ (green curve), $E_{\textrm{Th}}=5 \Delta$ (blue curve), and $E_{\textrm{Th}}=10 \Delta$ (yellow curve). The shaded regions in (b) correspond to variations of the energies with the position of the scatterer along the trajectory (described by the parameter $x$).}
\end{figure}

The scattering matrices for electron-hole conversion at the interface to the superconductors  are given by
$S_A^{he}(E,\varphi)=\exp[-i \arccos(E/\Delta)-i\varphi/2 \sigma_3]$
and $S_A^{eh}(E,\varphi)=S_A^{he}(E,-\varphi)$, respectively. 
Note that the $\sigma$-space is not Nambu space. An electron arriving at either superconductor is reflected as a hole traveling towards the normal region from the same side, thus Andreev reflection is described by a diagonal matrix. The condition for a bound state  reads as
\begin{equation}
	\det\left[1-S_N^e(E,x) S_A^{eh}(E,\varphi/2)S_N^h(E,x)S_A^{he}(E,-\varphi/2)\right]=0.
\end{equation}
The bound-state energies in dependence of $\varphi$ are plotted in Fig.~\ref{fig:12} for different values of $E_\textrm{Th}$. Without backscattering in the normal region, at $\varphi=0$ the two Andreev levels are degenerate [Fig.~\ref{fig:12} (a)]. Taking into account impurity scattering in the normal part [Fig.~\ref{fig:12} (b)] this degeneracy is lifted (the exact curve depends on the position where scattering occurs, i. e., on the parameter $x$). This results in the characteristic shape of the minigap and the secondary "smile"-gap below $E=\Delta$. Figure~\ref{fig:12}(b) shows the $x$-averaged results for Andreev bound states with one scattering event with $T=0.9$ (weak scattering). It is worth mentioning that only channels without scattering contribute to the zero-energy Andreev states at $\varphi=\pm \pi$ (not shown). For paths with one or more scattering events (more scattering matrices in the normal part), these levels are shifted to higher energies. Thus, we have shown that the secondary gap can be understood from the phase-dependence of the Andreev level when the junction length exceeds a length of the order of the superconducting coherence length, given by $E_\textrm{Th}\gtrsim\Delta$. The "smile" shape can be traced back to the effect of backscattering.

\section{Conclusion}

To summarize, we have calculated the local density of states for diffusive Josephson systems for a wide range of contact types with attention to the energy range below $\Delta$, in which a secondary gap can appear. We have generalized previous calculations for ballistic contacts \cite{reutlinger:14} and shown that the secondary "smile"-gap is a robust feature in the proximity density of states for large Thouless energies. We thus suggest that this feature should be accessible to an experimental detection by means of high-resolution scanning tunneling spectroscopy and want to encourage research in this direction. 

{\it Acknowledgments}.
J.~R. and W.~B. were supported by the DFG through SFB 767 and BE 3803/5 and by the Carl Zeiss Foundation. Y.~N. and L.~G. thank  the Aspen Center for Physics, supported in part by NSF Grant No. PHYS-1066293, for hospitality. Work at Yale is supported by NSF DMR Grant No. 1206612.


\begin{thebibliography}{9}

\bibitem{deutscher:69}
	G. Deutscher and P. G. de Gennes, in Superconductivity, edited
	by R. D. Parks (Dekker, New York, 1969), Vol. 2, p. 1005.
	
%
\bibitem{mcmillan:68}
	 W.~L. McMillan, Phys. Rev. \textbf{175}, 537 (1968). 

\bibitem{golubov:88} A.~A. Golubov and M.~Yu. Kuprianov, J. Low Temp. Phys. \textbf{70}, 83 (1988).

\bibitem{beenakker:92}
	C.~W.~J. Beenakker and H. van Houten, 
	in: \textit{Single-Electron Tunneling and Mesoscopic Devices}, 
	ed. by H. Koch and H. L\"ubbig (Springer, Berlin, 1992)

\bibitem{belzig:96} 
	W. Belzig, C. Bruder, and G. Sch\"on, 
	Phys. Rev. B \textbf{54}, 9443 (1996).

\bibitem{gueron:96}
	S. Gueron, H. Pothier, N.~O. Birge, D. Esteve, and M.~H. Devoret,
	Phys. Rev. Lett. \textbf{77}, 3025 (1996).

\bibitem{scheer:01}
	E. Scheer, W. Belzig,Y. Naveh, M.~H. Devoret, D. Esteve, and C. Urbina,
	Phys. Rev. Lett. \textbf{86}, 284 (2001).

\bibitem{moussy:07}
	N. Moussy, H. Courtois, and B. Pannetier,
	Europhys. Lett. \textbf{55}, 861 (2007)

\bibitem{mourik:12}
	V. Mourik, K. Zuo, S.~M. Frolov, S.~R. Plissard, E.~P.~A.~M. Bakkers, and L.~P. Kouwenhoven, 
	Science \textbf{336}, 1003 (2012).

\bibitem{churchill:13}
	H.~O.~H. Churchill, V. Fatemi, K. Grove-Rasmussen, M.~T. Deng, P. Caroff, H.~Q. Xu, and C.~M. Marcus, 
	Phys. Rev. B \textbf{87}, 241401 (2013).
	
\bibitem{garcia:13}
	L.~Serrier-Garcia,  J.~C. Cuevas, T. Cren, C. Brun, V. Cherkez,  F. Debontridder, D. Fokin, F.~S. Bergeret and D. Roditchev,  
	Phys. Rev. Lett. \textbf{110}, 157003 (2013)
	
\bibitem{cherkez:14}
	V. Cherkez, J.~C. Cuevas, C. Brun, T. Cren, G. Menard, F. Debontridder, V.~S. Stolyarov, and D. Roditchev,
	Phys. Rev. X \textbf{4}, 011033 (2014)	

\bibitem{leseur:08}
	H. le Sueur, P. Joyez, H. Pothier, C. Urbina, and D.  Esteve, 	
	Phys. Rev. Lett. \textbf{100}, 197002 (2008).
	
\bibitem{pillet:10}
	J.-D. Pillet, C. H. L. Quay, P. Morfin, C. Bena, A. Levy Yeyati, and P. Joyez, 
	Nat. Phys. \textbf{6}, 965 (2010).
	
\bibitem{wolz:11}
	M. Wolz, C. Debuschewitz, W. Belzig, and E. Scheer,
	Phys. Rev. B \textbf{84}, 104516 (2011).

\bibitem{josephson:62}
	B. D. Josephson, Phys. Letters \textbf{1}, 251 (1962).

\bibitem{belzig:99}  W. Belzig, F. K. Wilhelm, C. Bruder, G. Sch\"on, and A. D. Zaikin, 
	Superlatt. Microstruct. \textbf{25}, 1251 (1999).

\bibitem{golubov:04}
	A.~A. Golubov, M.~Yu. Kupriyanov, and E. Ilichev, 
	Rev. Mod. Phys. \textbf{76}, 411 (2004).

\bibitem{lodder:98}
	A. Lodder and Yu.~V. Nazarov, 
	Phys. Rev. B \textbf{58}, 5783 (1998).
	
\bibitem{pilgram:00} 
	S. Pilgram, W. Belzig, and C. Bruder, 
	Phys. Rev. B \textbf{62}, 12462 (2000).


\bibitem{VavilovLarkin:03} 
	M.~G. Vavilov and A.~I. Larkin,
	Phys. Rev. B \textbf{67}, 115335 (2003).

\bibitem{beenakker:05}
	C.~W.J~. Beenakker,
	Lect. Notes Phys. \textbf{667}, 131 (2005).

\bibitem{melsen:97} 
	J.~A. Melsen, P.~W. Brouwer, K.~M. Frahm, and C.~W.~J. Beenakker, 
	Physica Scripta T69, 233 (1997)

\bibitem{kuipers:10}
	J. Kuipers, D. Waltner, C. Petitjean, G. Berkolaiko, and K. Richter, 
	Phys. Rev. Lett. \textbf{104}, 027001 (2010).	
	
\bibitem{kuipers:11} 
	J. Kuipers, T. Engl, G. Berkolaiko, C. Petitjean,  D. Waltner, and K. Richter, 
	Phys. Rev. B {\bf 83} , 195316 (2011).

\bibitem{levchenko:08}	
	A. Levchenko, Phys. Rev. B \textbf{77}, 180503(R) (2008).

\bibitem{hammer:07} 
	J. C. Hammer, J. C. Cuevas, F. S. Bergeret and W. Belzig, 
	Phys. Rev. B \textbf{76}, 064514 (2007).

\bibitem{heikkilae:02} 
	T. T. Heikkilae, J. Särkkä and F. K. Wilhelm, 
	Phys. Rev. B \textbf{66}, 184513 (2002).

\bibitem{wilhelm:00} 
	F. K. Wilhelm and A. A. Golubov, 
	Phys. Rev. B \textbf{62}, 5353 (2000).

\bibitem{golubov:96}
	A. Golubov and M. Kupriyanov, Physica C \textbf{259}, 27 (1996).
	
\bibitem{bezuglyi:05} 
	E.~V. Bezuglyi, A.~S. Vasenko, V.~S. Shumeiko, G. Wendin,
	Phys. Rev. B \textbf{72}, 014501 (2005)
	
\bibitem{reutlinger:14} 
	J. Reutlinger, L. Glazman, Yu. V. Nazarov and W. Belzig
	Phys. Rev. Lett. \textbf{112}, 067001 (2014).

\bibitem{nazarov:94}
	Yu.~V. Nazarov, Phys. Rev. Lett. \textbf{73}, 134 (1994).

\bibitem{qt}
	Yu.~V. Nazarov and Ya. M. Blanter, \textit{Quantum Transport} (Cambridge University Press, Cambridge, 2009)

\bibitem{nazarov:99}
	Yu.~V. Nazarov, Superlatt. Microstruct. \textbf{25}, 1221 (1999).
	
\bibitem{golubov:89} A.~A. Golubov and M.~Yu. Kupriyanov, 
	ZhETF \textbf{96}, 1420 (1989) [Sov. Phys. JETP \textbf{69}, 805 (1989)].	
	
\bibitem{schep:97}
	K.~M. Schep and G. E. W. Bauer, Phys. Rev. Lett. \textbf{78}, 3015 (1997);
	Phys. Rev. B.  \textbf{56}, 15860 (1997).

\bibitem{dorokhov:82}
	O.~N. Dorokhov, 
	Solid State Commun. \textbf{51}, 381 (1984). 

\bibitem{belzig:00}
	W. Belzig, A. Brataas, Yu. V. Nazarov, G. E. W. Bauer,
	Phys. Rev. B \textbf{62}, 9726 (2000)
	
	
\end{thebibliography}

\end{document}